\documentclass[nonacm,sigconf,screen,balance=false]{acmart}

\makeatletter
\def\@ACM@checkaffil{}
\makeatother

\usepackage{amsmath,amsfonts}
\usepackage{listings}

\usepackage{amssymb}
\usepackage{graphicx}
\usepackage{xcolor}

\usepackage{pgfplots}
\usepgfplotslibrary{groupplots,fillbetween}
\pgfplotsset{compat=1.18}

\usepackage{lipsum}
\usepackage{booktabs,tabularx,tabulary} 
\usepackage{multirow,makecell}
\usepackage{placeins}
\usepackage{adjustbox}
\usepackage[table]{xcolor}
\usepackage{tikz}
\usepackage[most]{tcolorbox}
\usepackage{textcomp}
\usepackage{pgfplots}
\usepackage{hyperref}
\pgfplotsset{compat=1.18}
\usepackage{verbatim}
\usepackage{fancyvrb}
\usepackage[ruled,vlined,linesnumbered]{algorithm2e}
\usepackage{balance}

\definecolor{OpenBG}{RGB}{248,250,255}   
\definecolor{ClosedBG}{RGB}{255,249,240} 
\definecolor{RoleCircle}{RGB}{97,198,138}

\newtheorem{example}{Example}

\newcommand{\closedrow}{\rowcolor{ClosedBG}}
\newcommand{\openrow}{\rowcolor{OpenBG}}
\newcommand{\legendbox}[2]{%
  \begingroup
  \setlength{\fboxsep}{2pt}%
  \colorbox{#1}{\strut #2}%
  \endgroup
}

\newcommand{\instancecolorbox}[2]{%
  \begingroup
  \setlength{\fboxsep}{0pt}%
  \colorbox{#1}{\strut #2}%
  \endgroup
}

\newcommand{\legendtri}[2][red]{%
  \tikz[baseline=-0.5ex]\fill[#1] (0,0)--(#2,0)--(.5*#2,.85*#2)--cycle;%
}
\newcommand{\legendcirc}[2][RoleCircle]{%
  \tikz[baseline=-0.5ex]\fill[#1] (0,0) circle (.5*#2);%
}

\newcommand{\ignore}[1]{}

\hypersetup{
	pdfstartview=FitH,
	bookmarksnumbered=true,
	bookmarksopen=true,
	colorlinks,
	linkcolor={blue!90!black},
	citecolor={blue!60!black},
	urlcolor={blue!80!black}
}

\usepackage{enumitem}

\lstdefinestyle{mystyle}{
    commentstyle=\color{green},
    keywordstyle=\color{blue},
    stringstyle=\color{purple},
    basicstyle=\small\ttfamily,
    breaklines=true,
    columns=fullflexible,
    frame=single,
}

\usepackage{mdframed}

\begin{document}

\title{Benchmarking Text-to-SQL under Role-Based Access Control}

\author{Yang Fei}
\affiliation{
  \institution{National University of Singapore}
}
\email{yfei11@u.nus.edu}

\author{Yangfan Jiang}
\affiliation{
  \institution{National University of Singapore}
}
\email{jyangfan@u.nus.edu}
\author{Yin Yang}
\affiliation{
  \institution{Hamad Bin Khalifa University}
}
\email{yyang@hbku.edu.qa}

\author{Xiaokui Xiao}
\affiliation{
  \institution{National University of Singapore}
}
\email{xkxiao@nus.edu.sg}


\begin{abstract}
Given a database $\mathcal{S}$ and a natural language question $Q$, text-to-SQL systems aim to generate an SQL query that correctly answers $Q$ when executed against $\mathcal{S}$. Currently, popular text-to-SQL benchmarks mostly assume unrestricted access to $\mathcal{S}$; in practice, however, user access is often restricted, e.g., through \emph{role-based access control} (RBAC) policies. This leads to a potential disconnect between benchmarking results and real-world performance: an LLM with high benchmark scores might perform poorly in an access-controlled environment, by frequently violating RBAC, or rejecting a query $q$ that could be answered with only permitted data in $\mathcal{S}$.

Motivated by this, we present a comprehensive text-to-SQL benchmarking framework with realistic RBAC constraints, which features an LLM-assisted workflow that augments existing text-to-SQL benchmarks with plausible user roles and access policies. To do so, we formulate the problem of role synthesis as a structured reasoning process over the database schema, in which the LLM first infers the application context from the schema, and then derives role responsibilities and access scopes consistent with this context. This process is audited by human-in-the-loop quality control, in which domain experts perform metric-guided screening on the generated roles. Besides the augmented dataset, the proposed framework also contains evaluation metrics that identify RBAC-specific failure modes, and disentangle SQL utility from access-control compliance.

We apply the proposed framework to several widely-used benchmarks, and conduct a systematic empirical study of state-of-the-art text-to-SQL systems. The results show that many solutions (especially open-weight LLMs) with high benchmarking scores under an unrestricted setting suffer sharp performance degradation once access constraints are in place, due to frequent RBAC violations.

\end{abstract}

\maketitle

\section{Introduction}\label{sec:introduction}

Structured query language (SQL) has long served as the \textit{lingua franca} for database interactions. 
However, for many users, the complexity of SQL remains a significant barrier to effective data access~\cite{li2014nalir,kim2020natural}. 
Text-to-SQL systems aim to eliminate this barrier by allowing users to express their intent in natural language and translating it into SQL. 
Recent advances in large language models (LLMs), with strong coding and reasoning capabilities, have enabled substantial progress in text-to-SQL performance~\cite{li_can_2023,gao_text--sql_2024,luoma2025snails,xie2025opensearch,chen2025reliable}.

State-of-the-art LLM-based text-to-SQL solutions typically combine prompt engineering and schema-aware reasoning to achieve high execution accuracy~\cite{zhou_db-gpt-hub_2024}. 
To evaluate and compare different approaches, several benchmarks have been established, including Spider~\cite{yu_spider_2019,lei_spider_2025}, BIRD~\cite{li_can_2023}, and LiveSQLBench~\cite{livesqlbench_hf}. 
On these benchmarks, recent systems report strong performance under the assumption that generated SQL queries are executed directly by the DBMS without additional constraints.

\subsection{Text-to-SQL in the Wild}\label{subsec:motivation}
In real-world DBMS environments, database access is rarely unrestricted. 
To satisfy security, privacy, and regulatory requirements, production DBMSs enforce \emph{access control policies} that determine which data a user is authorized to access. 
One of the most widely adopted mechanisms is {role-based access control} (RBAC), 
under which users are assigned roles with permissions over specific tables, attributes, operations, or views~\cite{485845, ferraiolo2003role}. 
In such settings, an unauthorized SQL query could trigger an execution error at the DBMS layer (e.g., MySQL's error 1142\footnote{\url{https://dev.mysql.com/doc/refman/9.5/en/grant.html}}), or silent result suppressions (e.g., Snowflake's dynamic data masking\footnote{\url{https://docs.snowflake.com/en/user-guide/security-column-ddm-intro}}) that are non-trivial to debug.

This exposes a fundamental gap between real-world deployments and current text-to-SQL evaluation practices. 
RBAC enforcement should remain deterministic at the DBMS layer; however, modern text-to-SQL systems typically generate SQL outside the DBMS and must therefore make access-sensitive decisions before execution.
In practice, this separation can lead to repeated execution failures, brittle system behavior, and subtle mismatches between user intent and the query results returned to the user, often requiring costly manual intervention.
A DBMS-side checker can reject unauthorized SQL at execution time, but by that point the upstream text-to-SQL workflow may already have failed to produce a usable response.

To illustrate this issue, consider the following example.

\begin{example}\label{example:rbac-sql}
Consider an enterprise analytics application in which a text-to-SQL  framework translates user requests into SQL queries executed by a backend DBMS. 
Suppose a user with the ``Sales Analyst'' role issues the natural language query:
``Which departments have the highest average sales revenue per employee?''
A vanilla text-to-SQL system may generate the following SQL query:

\lstset{style=mystyle}
\begin{lstlisting}[language=SQL,
deletekeywords={IDENTITY},
deletekeywords={[2]INT},
morekeywords={clustered},
mathescape=true,
xleftmargin=20pt,
framexleftmargin=0pt,
frame=tb,
framerule=0pt]
SELECT e.department, AVG(s.amount) AS avg_revenue
FROM employees e
JOIN sales s ON e.emp_id = s.emp_id
GROUP BY e.department;
\end{lstlisting}

Under a typical RBAC policy, however, the ``Sales Analyst'' role may be permitted to access the \texttt{sales} table but not sensitive employee information such as identifiers or department assignments.
Although the generated query is syntactically valid and semantically aligned with the user's intent, it violates the RBAC policy and is rejected by the DBMS at execution time. 
As a result, even if a deterministic checker blocks the query, the application-level workflow fails to complete: the text-to-SQL component neither produces an authorized query nor correctly refuses the request before execution. 
The system is therefore safe at the DBMS layer but not useful at the application layer, leaving the request unresolved and requiring retries or manual intervention.
More generally, such a checker only determines whether a generated SQL query is authorized; it does not reveal whether the system should have refused earlier, whether it missed a policy-compliant SQL alternative, or whether it over-refused an answerable request.
\end{example}

Despite the practical importance of access control, most existing text-to-SQL benchmarks still focus on translation accuracy under unrestricted database access.
{
While a few recent efforts~\cite{weng_bridgescope_2025,subramaniam_deploi_2025,klisura2025role} have begun to consider access control, they mainly target coarse-grained settings (e.g., database- or table-level permissions) and lack systematic benchmarking pipelines or comprehensive empirical analysis, offering limited insight into practical text-to-SQL deployments under RBAC.
}
To our knowledge, established benchmarks and evaluation protocols~\cite{yu_spider_2019, li_can_2023, livesqlbench_hf,ren2024purple,gao_text--sql_2024,zhang_finsql_2024,luoma2025snails,yang2025automated,xie2025opensearch,chen2025reliable,fan2024combining,fu2023catsql,chung25longcontextsql} do not account for RBAC.
This leaves an overlooked failure mode that we term an \emph{RBAC-rejected success}, 
where a generated SQL query is syntactically correct and would be judged correct under unrestricted-access evaluation, yet is rejected at execution time due to RBAC violations.
Since existing benchmarks assume unrestricted access, standard metrics such as execution accuracy (EX) fail to capture such failures.
As a result, current benchmarks and metrics do not measure a text-to-SQL system's ability to generate {authorized} SQL queries under realistic access control constraints, leaving a practically important class of errors largely invisible.

As we shall demonstrate in Section~\ref{sec:experiments}, RBAC-rejected successes are in fact common. Systems that achieve high EX scores on existing benchmarks may frequently fail once RBAC rules are enforced, 
since simply prompting LLMs with access policies does not reliably prevent unauthorized SQL generation. 
Moreover, heuristic remedies such as schema filtering, prompt engineering, and fine-tuning are insufficient; LLMs may still hallucinate relationships or make flawed assumptions, leading to invalid SQL and failed executions.
These observations indicate that RBAC alignment must be treated as a first-class concern in text-to-SQL systems rather than an afterthought handled solely at execution time. 
Accordingly, a critical mismatch remains between the strong performance reported on existing benchmarks and the requirements of practical, access-controlled deployments in real DBMS environments.

\subsection{Contributions}\label{subsec:contributions}
\begin{figure*}[!t]
    \centering
    \includegraphics[width=1.0\linewidth]{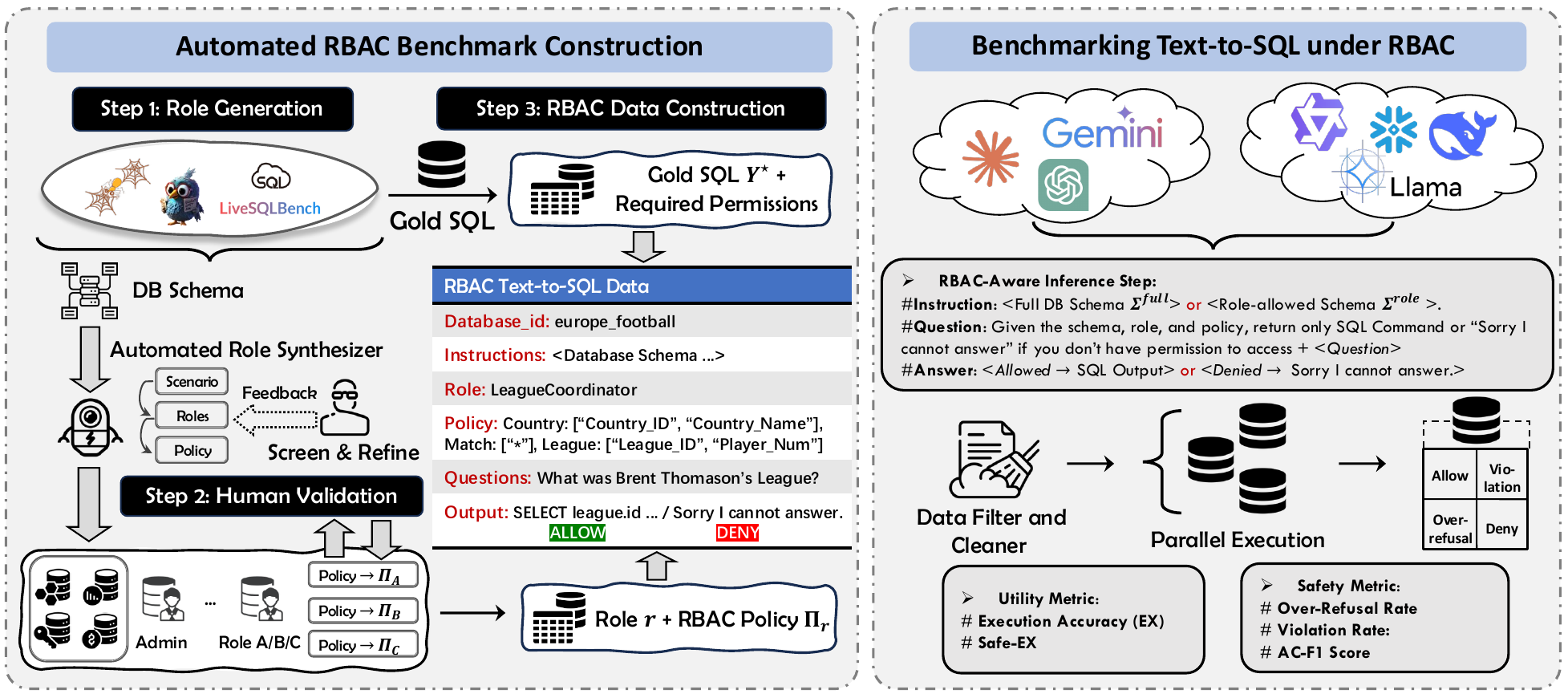}
    \vspace{-4.5mm}
    \caption{Overview of the proposed RBAC-aware dataset construction and benchmarking pipeline.}
    \label{fig:overview}
    \vspace{-1mm}
\end{figure*}

To fill this gap, we present a comprehensive benchmarking framework for evaluating the RBAC-enforcing capabilities of text-to-SQL systems, as illustrated in Figure~\ref{fig:overview}. 
In a nutshell, the proposed framework takes as input an existing text-to-SQL dataset, augments the data with plausible roles and access policies generated with an LLM with human-in-the-loop validation, and defines new evaluation metrics that evaluate both SQL correctness and RBAC compliance.
We apply this framework to three widely-used benchmarks, namely Spider~\cite{yu_spider_2019,lei_spider_2025}, BIRD~\cite{li_can_2023}, and LiveSQLBench~\cite{livesqlbench_hf}, producing RBAC-augmented datasets that support the evaluation of both access-policy compliance and SQL utility, and conduct a systematic empirical study that reveals previously unexamined behaviors of text-to-SQL systems under access control. 

Our goal is not to replace deterministic DBMS-side enforcement with an LLM, but to evaluate the upstream text-to-SQL problem.
We study whether a system can make the correct allow/deny decision given the natural language query and role policy and, when allowed, generate correct, policy-compliant SQL.
A DBMS-side checker remains complementary, but does not resolve this evaluation problem.
It can reject unauthorized SQL, yet does not indicate whether the system should have refused earlier, generated a policy-compliant alternative, or how much utility is lost to over-refusal.

\vspace{1mm}
\noindent\textbf{Challenges.}
Designing an RBAC-aware text-to-SQL benchmark faces four challenges.
First, synthesized roles and fine-grained policies must be semantically meaningful: they should reflect plausible application contexts while avoiding trivial or redundant permission patterns.
Second, reliable ground truth requires precise permission verification, as each reference SQL must be analyzed to identify all required permissions over tables, columns, and operations.
Third, automatically generated roles and policies may be unrealistic or imbalanced, requiring human validation of semantic plausibility, permission coverage, and role distinctness.
Finally, existing text-to-SQL metrics do not capture RBAC-specific failures, 
motivating new metrics that jointly measure SQL utility and policy compliance.

\vspace{1mm}
\noindent\textbf{Framework overview.}
At the core of the proposed framework is an automated and auditable pipeline that augments existing text-to-SQL benchmarks with realistic RBAC settings while preserving their original linguistic diversity and query semantics.
For each database, the pipeline uses LLMs to synthesize plausible user roles and fine-grained access policies.
To avoid degenerate role definitions, role synthesis follows a structured reasoning process tailored to text-to-SQL under RBAC constraints. Specifically, the model first infers an application context from the database schema, and then derives role responsibilities and access policies consistent with the schema structure.
The resulting policies operate at the column-operation level, enabling evaluation under restricted access.

We further incorporate \textit{human-in-the-loop} validation to ensure that the synthesized roles and policies are meaningful and faithful to practical settings.
LLM-generated role configurations are first screened using interpretable quality metrics, including denial rate, coverage balance across operations, role distinctness, and semantic alignment with the database schema, and are then reviewed by human evaluators for plausibility.
Configurations that fail these checks are rejected, with failure signals fed back to the synthesis stage to refine the role definitions and access policies.
For permission assignment, each reference SQL query is parsed using SQLGlot~\cite{sqlglot} to extract the required permissions, which are then validated against each role's access policy.
This process produces RBAC-aware ground truth that distinguishes between allowed queries, proper refusals, and unsafe cases in which a semantically correct query would violate access constraints.

Finally, we introduce evaluation metrics that explicitly disentangle SQL utility from policy compliance.
These metrics capture not only whether a generated query produces the correct result under full access, but also whether it respects user permissions and avoids RBAC-rejected successes, namely cases in which syntactically correct SQL violates access policies.

\vspace{1mm}
\noindent\textbf{Benchmarking results and analysis.}
We apply this framework to several widely used text-to-SQL benchmarks, resulting in large-scale evaluation resources spanning 53 databases, 399 tables, and 3,353 columns, with a total of 21,502 RBAC-annotated query instances.
Using these benchmarks, we conduct a systematic empirical study of state-of-the-art text-to-SQL systems and observe persistent RBAC-related failures.
Failure analysis reveals the \textit{refusal-cliff} phenomenon, i.e., an LLM may correctly follow RBAC rules during reasoning, yet still output a final SQL query that violates access control policies.
Meanwhile, common mitigation measures, including \textit{supervised fine-tuning} and \textit{in-context learning}, show limited effectiveness in reducing RBAC failures.
These findings highlight a substantial gap between strong benchmark performance under unrestricted settings and the requirements of access-controlled deployments, motivating text-to-SQL systems that explicitly reason about RBAC constraints.

We release the complete benchmarking pipeline, evaluation toolkit, and the
resulting RBAC-augmented datasets in a public repository~\cite{yang2026rbactext2sql}, to support reproducibility and future research.

\section{Problem Setting}\label{sec:preliminaries}
This section formalizes the RBAC-compliant text-to-SQL task studied in this paper, including its objectives and schema exposure mechanisms. We then describe the LLM-based role (profile) synthesis problem used to construct our benchmark. 

\subsection{Text-to-SQL under RBAC}\label{sec:pre:text-to-sql}

\noindent
\textbf{LLM-based text-to-SQL.} 
Let $\mathcal{S}=(\mathcal{T},\mathcal{C},\mathcal{R})$ denote a relational database schema, where $\mathcal{T}$ is the set of tables, $\mathcal{C}$ the set of columns, and $\mathcal{R}$ the set of foreign-key relations.
Given a natural-language question $Q$ and a schema text $\Sigma$ describing $\mathcal{S}$, the goal of text-to-SQL is to generate a SQL program $Y$ whose execution result matches that of a canonical gold standard query $Y^\star$.

An LLM-based system forms a prompt $\mathcal{P}(Q,\Sigma)$ and decodes an SQL program token by token using a language model $M$:
\[
P_M(Y \mid \mathcal{P}(Q,\Sigma))=  \prod_{i=1}^{|Y|} P_M\!\left(y_i \,\middle|\, \mathcal{P}(Q,\Sigma),\, y_{<i}\right).
\]
Through the model $M$, a candidate program $\hat{Y}$ is obtained by greedy or sampling-based decoding, and then executed for evaluation.

\vspace{1mm}
\noindent
\textbf{Text-to-SQL with RBAC.} 
We extend the standard text-to-SQL setting by introducing role-based access control.
Each role $r$ is associated with an access policy
$\Pi_r \subseteq \mathcal{T} \times \mathcal{C} \times \mathcal{O}$,
where $\mathcal{O}$ denotes the set of SQL operations (e.g., \texttt{SELECT}, \texttt{INSERT}, \texttt{UPDATE}, \texttt{DELETE}).
A permission $(t,c,o)\in\Pi_r$ authorizes role $r$ to apply operation $o$ to column $c$ of table $t$. Given a gold SQL $Y^\star$, we extract the required permission set 
$\mathsf{Perm}(Y^\star) \subseteq \mathcal{T} \times \mathcal{C} \times \mathcal{O}$,
corresponding to all $(\mathsf{table}, \mathsf{column}, \mathsf{operation})$ triples exercised by $Y^\star$. The ground-truth access decision for an instance $(Q,\mathcal{S},r)$ is defined as:
\[
y \;=\;
\begin{cases}
\textsf{allow}, & \text{if }\; \mathsf{Perm}(Y^\star)\subseteq \Pi_r,\\[2pt]
\textsf{deny},  & \text{otherwise}.
\end{cases}
\]
This construction follows the standard assumption adopted in text-to-SQL benchmarks that the gold SQL, authored by domain experts, accurately captures the data access and operations required to answer the query.
Accordingly, in our benchmark, we treat the gold SQL as a correct and complete operational specification of the user's intended data access behavior.

A system receives $(Q,\Sigma,r,\Pi_r)$, where $\Pi_r$ is the serialized role policy provided in the prompt, and returns either a refusal symbol $\bot$ or a SQL program $\hat{Y}$. 
Emitting $\bot$ induces $\hat{y}=\textsf{deny}$; emitting $\hat{Y}$ induces $\hat{y}=\textsf{allow}$, after which $\hat{Y}$ is evaluated along two dimensions:
(i) \emph{RBAC compliance}, by checking whether $\mathrm{Perm}(\hat{Y}) \subseteq \Pi_r$, and
(ii) \emph{execution correctness}, by comparing its execution result with that of the gold SQL $Y^\star$.
Notably, even when $Y^\star$ is authorized under the policy $\Pi_r$, the generated SQL $\hat{Y}$ may still violate RBAC by invoking unnecessary or unauthorized tables, columns, or operations.

We consider two modes for the schema available: (i) $\Sigma^{\mathrm{full}}$, which describes all information in $\mathcal{S}$; (ii) $\Sigma^{\mathrm{role}}(r)$, which describes only schema elements permitted by $\Pi_r$. 
This controls the information exposed at prompt time without changing the underlying database. 

\vspace{1mm}
\noindent
\textbf{Task objective.}
For each instance, the model must produce the correct access decision $y$ under the role's RBAC policy. 
Conditional on both the ground-truth and predicted decision being \textsf{allow}, the model must further generate a SQL program $\hat{Y}$ that is execution-correct with respect to $Y^\star$ and compliant with the role's fine-grained permissions.
This formulation naturally induces a two-stage task structure, separating policy compliance from SQL generation quality while remaining compatible with execution-based evaluation.

\vspace{1mm}
\noindent\textbf{Scope and limitations.}
Our benchmark focuses on \emph{explicit} RBAC compliance: whether the system output references only the tables, columns, and operations authorized by the role policy~$\Pi_r$.
It does not model \emph{inference-based} leakage, such as inferring salary ranges from authorized columns \texttt{hourly\_wage} and \texttt{hours\_worked} when \texttt{salary} is denied.
Such indirect leakage requires policy models beyond standard RBAC, such as inference control or semantic privacy, and is left as future work in Section~\ref{sec:conclusion}.

\subsection{RBAC Role Synthesis}\label{sec:profile}
To construct a realistic benchmark for RBAC-compliant text-to-SQL, an important task is to systematically synthesize a set of plausible role profiles for a given database schema $\mathcal{S}$. 
We refer to this process as \emph{role synthesis}.
Given $\mathcal{S}$, the objective is to generate a role set $\mathbf{R}_{\mathcal{S}}=\{r_1,r_2,\dots,r_k\}$,
where each role $r_i$ is defined as a tuple $(n_i,d_i,\Pi_{r_i})$ consisting of a role name $n_i$, a natural-language description $d_i$, and a fine-grained RBAC policy $\Pi_{r_i}$. 

\begin{table}[t]
\centering\small
\caption{Databases, tables, columns, and query counts per data source with and without role design.}
\vspace{-2mm}
\centering
\small
\setlength{\tabcolsep}{4pt}
\begin{tabular}{l r r r r r}
\toprule
\textbf{Data Source} & \textbf{DBs} & \textbf{Tables} & \textbf{Columns} & \textbf{w/o role} & \textbf{w/ role} \\
\midrule
Spider        & 20 & 80 & 439 & 1,034 & 6,926 \\
BIRD          & 11 & 75 & 798 & 1,534 & 10,175 \\
LiveSQLBench  & 22 & 244 & 2,116 & 592 & 4,401 \\
\midrule
\textbf{Total} & \textbf{53} & \textbf{399} & \textbf{3,353} & \textbf{3,160} & \textbf{21,502} \\
\bottomrule
\end{tabular}
\vspace{-2mm}
\label{tab:query_counts}
\end{table}

An ideal role set should satisfy the following key properties.
First, the synthesized roles must be semantically plausible and reflect real-world job functions and responsibilities that are logically grounded in the application domain of the database $\mathcal{S}$.
For example, in an enterprise data analytics setting such as in Example~\ref{example:rbac-sql}, roles such as \textit{Sales Analyst} and \textit{HR Admin} are coherent, whereas \textit{Student} and \textit{Teacher} are not. 
Second, the roles should represent a reasonable division of responsibilities. The objective is not to create a strict partition (e.g., creating a separate role for 
every single table), as real-world roles often have overlapping permissions, but to avoid trivial or redundant role definitions. Ideally, 
generated roles should be semantically distinct, and reflect a realistic organizational structure.

Formally, automated role synthesis seeks to construct a role set $\mathbf{R}_{\mathcal{S}}$ that balances semantic plausibility and role distinctness. 
In our framework, this synthesis process is implemented using an LLM as a generative backend, and is realized through a structured pipeline that combines constrained generation with human-in-the-loop screening and iterative refinement, ensuring that the resulting role set $\mathbf{R}_{\mathcal{S}}$ satisfies the above criteria.

\section{RBAC-Aware Benchmark Construction}\label{sec:data_generation}
We design a three-step, automated pipeline for constructing RBAC-aware text-to-SQL datasets, illustrated in Figure~\ref{fig:overview}. 
The pipeline augments existing text-to-SQL benchmarks with role information and access-control semantics by leveraging state-of-the-art LLMs for role synthesis and structured analysis for permission verification. 
In what follows, Section~\ref{sec:data_gen:source} describes the source text-to-SQL datasets, Section~\ref{sec:data_gen:generation} presents our role generation framework, and Section~\ref{sec:data_gen:data_construction} details the RBAC dataset construction process.

\subsection{Data Sources and Pre-Processing}\label{sec:data_gen:source}
We ground our study on three well-established text-to-SQL corpora: 
(i) \textit{Spider}, a benchmark for complex cross-domain semantic parsing, equipped with a public evaluation framework and curated SQLite resources~\cite{yu_spider_2019}; 
(ii) \textit{BIRD}, which incorporates large-scale, real-world databases from diverse professional domains, thereby strengthening realism in data noise, external knowledge dependency, and query complexity~\cite{li_can_2023}; and (iii) \textit{LiveSQLBench-Base-Full-v1}, which simulates end-to-end workloads from enterprise practices, supplemented with auditable executing and testing scripts~\cite{livesqlbench_hf}.
For simplicity, hereafter we use the original names Spider, BIRD, and LiveSQLBench to denote the filtered source datasets.

To achieve a unified RBAC-aware evaluation while preserving the original design intents of the source datasets, we apply minimal and auditable preprocessing.
Specifically, we adopt the official development splits of Spider and BIRD to support reproducible experiments without relying on closed-source test servers.
For LiveSQLBench, we keep analytical queries together with  administrative management operations to broaden the task surface across all queries.
We also utilize the official database backend from each data source, SQLite execution for Spider and BIRD, and PostgreSQL for LiveSQLBench. We further develop a general-purpose evaluator that supports execution with values based on these backends. 
The resulting RBAC-aware evaluation pipeline preserves the original question statements, gold SQL answers, data content, and official database backends.
Table~\ref{tab:query_counts} reports the affected counts alongside corpus statistics.

\begin{figure*}[!t]
  \centering
  \includegraphics[width=1.02\textwidth]{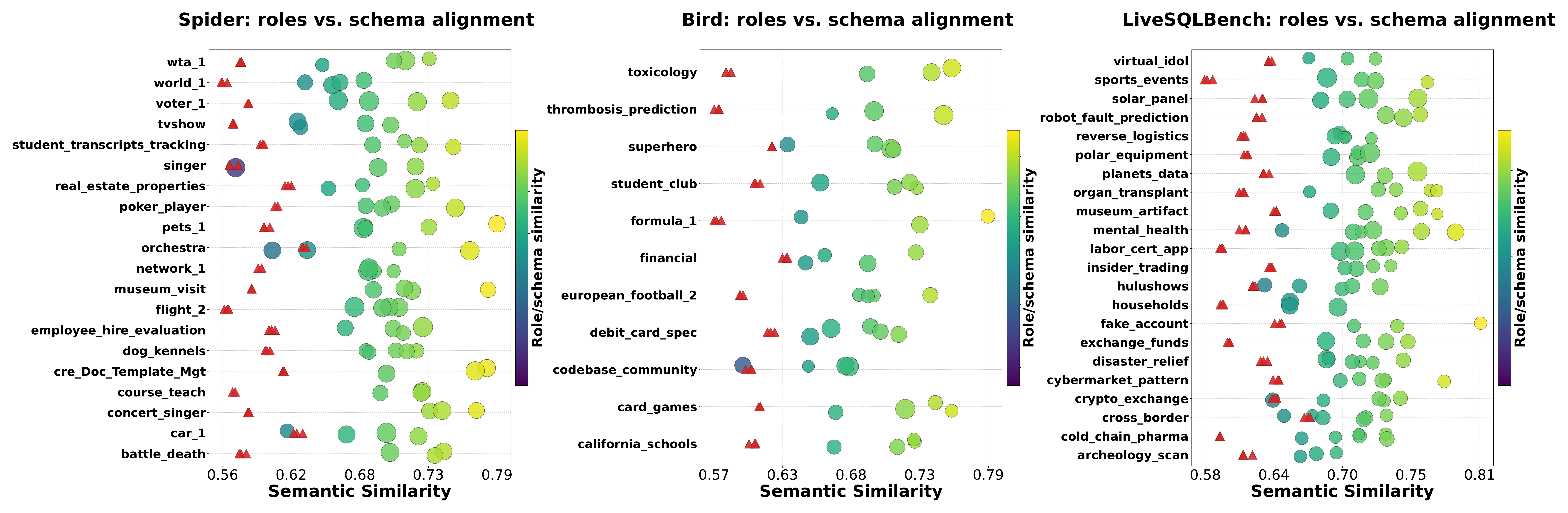}
  \vspace{-6.5mm}
  \caption{\textbf{Role Coverage and Semantic Alignment}. Each row (e.g., wta\_1) corresponds to a database. Each circular marker \legendcirc{1.6ex} represents a synthesized role, with size indicating the proportion of accessible columns and horizontal axis representing semantic similarity to the schema (purple: low, yellow: high). Red triangles \legendtri[red]{1.6ex} denote \textit{DataOperator} roles.}
  \label{fig:wide}
  \vspace{-1mm}
\end{figure*}

\subsection{Automated Role and Policy Synthesis}\label{sec:data_gen:generation} 
As described in Section \ref{sec:profile}, given a database 
$\mathcal{S}$, we synthesize a set of user roles $\mathbf{R}_{\mathcal{S}}$ using an automated role synthesis pipeline with an LLM 
and human-in-the-loop validation.
The LLM is provided with structured schema information, including table names, column attributes, data types, primary keys, and foreign key relationships, which together define the logical organization of the database.

A naive instantiation of this idea is to present the schema as plain text and directly prompt the model to generate roles.
In practice, however, this approach often produces incoherent or overly fragmented role sets that fail to capture implicit domain relationships and realistic access patterns. 
For example, consider a school database containing tables \textsf{StudentGrade}, \textsf{TeacherSalary}, and \textsf{CampusBuilding}. 
When prompted directly, the model may generate isolated roles such as \textit{Student} with access only to \textsf{StudentGrade}, \textit{Teacher} with access only to \textsf{TeacherSalary}, and \textit{Architect} with access only to \textsf{CampusBuilding}. 
While such roles may satisfy the basic schema coverage constraint in Section~\ref{sec:profile}, they fail to reflect realistic role responsibilities. 
In practice, teachers require access to student grades for assessment, and both students and teachers typically share access to campus facilities.

\vspace{1mm}
\noindent
\textbf{Context-aware role and policy construction.}
To address these limitations, we adopt a structured role synthesis strategy that constrains the model's inference process using schema semantics, rather than treating role generation as a free-form text completion task. 
Specifically, we decompose role synthesis and policy construction into a sequence of semantically grounded reasoning steps that guide the model to reason about the database at multiple levels of abstraction. 

As illustrated in Step 1 of Figure~\ref{fig:overview}, the model is first guided to infer a coherent application context implied by the database schema. This step requires the model to reason over table semantics, attribute names, and foreign key relationships to identify high-level usage scenarios, such as grading, enrollment management, or administrative workflows in an academic database. The inferred context serves as an intermediate semantic representation that anchors subsequent role construction. 
Conditioned on the inferred application context, the model then derives a small set of user roles together with their intended responsibilities. 
Each role is required to correspond to a meaningful organizational function, rather than an arbitrary subset of tables and columns. The synthesis process is explicitly structured to allow realistic overlap and dependency among roles, reflecting common access patterns in real DBMSs.

After synthesizing the role set, the model proceeds to derive a corresponding access control policy through a context-aware reasoning stage. 
Rather than assigning permissions independently, the model derives fine-grained access scopes that are consistent with the role's responsibilities, including column-level and operation-level permissions, depending on the target benchmark. 
By separating context inference, role responsibility definition, and permission derivation, this structured reasoning process reduces degenerate role definitions and encourages semantically coherent and internally consistent RBAC policies. 
Finally, we synthesize scoped administrator roles to better reflect practical DBMS deployments. Each database is assigned 2 or 3 \textit{DataOperator} roles with overlapping but incomplete access scopes. These roles provide broad schema coverage while still creating non-trivial deny cases for evaluation. Details are deferred to Appendix~\ref{appendix:gen-do}.

\vspace{1mm}
\noindent
\textbf{Automated screening and human validation.}
To ensure that the synthesized roles and their corresponding access policies are meaningful and faithful to practical settings, 
we incorporate a \textit{human-in-the-loop} validation stage into the role synthesis pipeline.
The validation process follows a reject-regenerate paradigm that combines automatic quality screening with targeted human inspection, while maintaining explicit records of validation signals and regeneration decisions for auditability purposes.

We first perform an automatic quality check on each synthesized role configuration using a set of interpretable metrics, including denial rate, coverage balance across query operations, role distinctness measured via permission overlap, and semantic alignment between role descriptions and the database schema; details of the metrics definitions are deferred to Appendix~\ref{appendix:human-eval}. 
These checks are fully automated and deterministic, producing explicit pass or fail signals together with diagnostic metrics that characterize the quality of the generated roles. 
When a role configuration fails any check, the specific failure reasons and associated metric diagnostics are recorded and packaged as structured feedback to the role synthesis stage. 
The context-grounded role and policy generation process is then re-executed using this feedback, enabling iterative refinement in a transparent and reproducible manner. 
The automatic reject-regenerate loop iterates until all quality checks are satisfied.

Once a role configuration passes all automatic checks, it is passed to human validation.
Four annotators with database and access-control expertise first complete a lightweight calibration on held-out cases to align the review criteria. They then review each synthesized role configuration, including its roles and policies, for semantic plausibility and consistency with the inferred application context.
Each configuration receives a binary accept/reject judgment and is accepted only if at least three of the four annotators approve it; otherwise, it is rejected and regenerated using the collected rejection reasons as structured feedback.
Only role configurations that pass both automatic screening and human validation are retained for subsequent permission verification and dataset construction.
Note that human validation is performed at the \emph{database-level} role configuration level (53 in total), not at the expanded role-query instance level. Once a configuration is accepted, allow/deny labels for all role-query pairs are produced automatically via structured SQL analysis and policy matching. Detailed validation statistics, including first-pass acceptance, regeneration,
manual revision, and total annotation time, are reported in Appendix~\ref{appendix:human-validate-stats}.

\vspace{1mm}
\noindent\textbf{Validation via semantic alignment.}
We further assess whether synthesized roles are semantically aligned with the underlying database schema.
For each role, we concatenate its name and description and embed the resulting text using a sentence embedding model (OpenAI/text-embedding-3-small in our experiments).
We similarly embed a schema description formed from the database's table and column names.
We compute cosine similarity between the role and schema embeddings as a lightweight proxy for semantic alignment.
For example, \emph{FinancialAuditor} should be closer to a database with \texttt{transactions} and \texttt{accounts} tables than to one with \texttt{players} and \texttt{matches}.
This check helps verify that LLM-generated roles are coherent with their assigned databases, rather than arbitrary.

Figure~\ref{fig:wide} reports the results on the three source datasets. 
Across all datasets, most synthesized roles show high semantic alignment with the schema while covering only a subset of the database access space. 
In contrast, \textit{DataOperator} roles show lower semantic similarity, consistent with their design as scoped administrators.
This pattern indicates that ordinary synthesized roles capture role-specific database semantics, while administrators mainly serve as structural safeguards rather than semantically specialized roles.

Human experts inspect the visualized results, identify outlier roles with low semantic similarity to the application context (e.g., the leftmost role in the \textit{singer} database of Spider), and provide feedback to the LLM for regeneration.

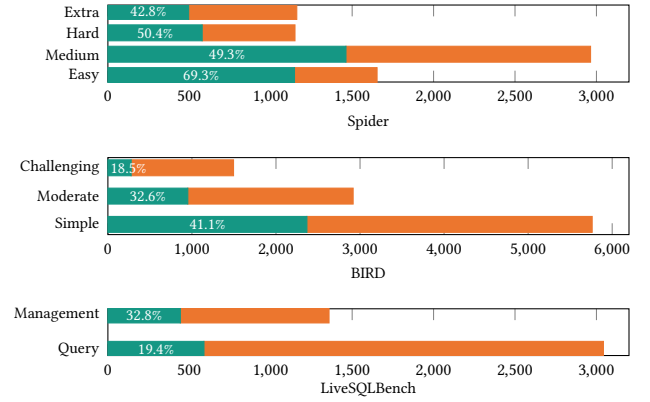
\begin{figure}[!tbp]
\centering
\begin{tikzpicture}
\definecolor{AllowGreen}{RGB}{21,151,132}
\definecolor{DenyOrange}{RGB}{236,118,46}

\begin{axis}[
  title={Spider},
  title style={font=\scriptsize, at={(0.5,-0.52)}, anchor=north},
  width=\columnwidth, height=2.6cm,
  xbar stacked, bar width=6pt,
  xmin=0, xmax=3200,
  symbolic y coords={Easy,Medium,Hard,Extra},
  ytick style={draw=none},
  ytick=data,
  yticklabel style={font=\scriptsize},
  xticklabel style={font=\scriptsize},
  label style={font=\scriptsize},
  legend to name=AllowDenyLegend,
  legend columns=2,
  legend style={/tikz/column sep=6pt, draw=none, fill=none, font=\scriptsize}
]
\addplot+[fill=AllowGreen, draw=AllowGreen,
          nodes near coords,
          point meta=explicit symbolic,
          every node near coord/.style={font=\scriptsize, text=white}]
  coordinates {(1146,Easy)   [69.3\%]
               (1462,Medium) [49.3\%]
               (579,Hard)    [50.4\%]
               (496,Extra)   [42.8\%]};
\addplot+[fill=DenyOrange, draw=DenyOrange]
  coordinates {(507,Easy)
               (1502,Medium)
               (570,Hard)
               (664,Extra)};
\legend{Allow, Deny}
\end{axis}

\begin{axis}[
  at={(0pt,-1cm)}, anchor=north west,
  width=\columnwidth, height=2.6cm,
  xbar stacked, bar width=6pt,
  xmin=0, xmax=6200,
  symbolic y coords={Simple,Moderate,Challenging},
  ytick style={draw=none},
  ytick=data,
  enlarge y limits=0.18,
  yticklabel style={font=\scriptsize},
  xticklabel style={font=\scriptsize},
  label style={font=\scriptsize},
  title={BIRD},
  title style={font=\scriptsize, at={(0.5,-0.52)}, anchor=north}
]
\addplot+[fill=AllowGreen, draw=AllowGreen,
          nodes near coords,
          point meta=explicit symbolic,
          every node near coord/.style={font=\scriptsize, text=white}]
  coordinates {(2368,Simple)      [41.1\%]
               (951,Moderate)     [32.6\%]
               (277,Challenging)  []}
  node[pos=0.97, anchor=center, xshift=-1pt, font=\scriptsize, text=white] {18.5\%};
\addplot+[fill=DenyOrange, draw=DenyOrange]
  coordinates {(3394,Simple)
               (1967,Moderate)
               (1218,Challenging)};
\end{axis}

\begin{axis}[
  at={(0pt,-3cm)}, anchor=north west,
  width=\columnwidth, height=2.2cm,
  xbar stacked, bar width=6pt,
  xmin=0, xmax=3200,
  symbolic y coords={Query,Management},
  ytick style={draw=none},
  ytick=data,
  enlarge y limits=0.18,
  yticklabel style={font=\scriptsize},
  xticklabel style={font=\scriptsize},
  label style={font=\scriptsize},
  title={LiveSQLBench},
  title style={font=\scriptsize, at={(0.5,-0.72)}, anchor=north}
]
\addplot+[fill=AllowGreen, draw=AllowGreen,
          nodes near coords,
          point meta=explicit symbolic,
          every node near coord/.style={font=\scriptsize, text=white}]
            coordinates {
              (590,Query)        [19.4\%]
              (445,Management)   [32.8\%]
            };
\addplot+[fill=DenyOrange, draw=DenyOrange]
            coordinates {
              (2453,Query)
              (913,Management)
            };
\end{axis}
\end{tikzpicture}
\vspace{-2mm}
\caption{\textbf{Query distribution and difficulty. Stacked bars show allowed (green) vs. denied (orange) query execution rates for each dataset and difficulty level/task category.}}
\label{fig:allow-deny-tikz}
\vspace{-3mm}
\end{figure}

\subsection{RBAC-Aware Instance Construction}\label{sec:data_gen:data_construction}
As shown in Figure~\ref{fig:overview}, after obtaining a validated role set $\mathbf{R_S}$ and the corresponding access policies for each database, the dataset construction phase expands each source text-to-SQL instance into role-conditioned evaluation instances. 
In particular, for each example $(\mathcal{S}, Q, Y^\star)$ in a source dataset, where $Q$ is the natural language question and $Y^\star$ is the gold SQL, the framework pairs it with every synthesized role $r \in \mathbf{R_S}$ associated with the same database. 
Each resulting instance inherits the original database identifier, question, and gold SQL, and is augmented with role-specific information, including the role description and the corresponding access policy.

A critical component of this phase is the permission verification mechanism, 
which deterministically labels each role-query pair as allowed or denied. 
We perform structured SQL analysis to extract the permissions required by the gold SQL $Y^\star$.
Specifically, we parse $Y^\star$ using SQLGlot~\cite{sqlglot} to deterministically recover the set of referenced base tables and accessed columns, and map $Y^\star$ to its CRUD operation type.
To obtain a complete and accurate dependency set, we analyze clause constructs like \texttt{JOIN}, \texttt{GROUP BY}, \texttt{ORDER BY}, and predicate filters, which may introduce additional column references in a complex SQL.
We then validate these extracted requirements against the role's access policy to decide whether the query should be allowed.
The extracted requirement set is then matched against the role's fine-grained access policy, which may specify permissions at the column level as per operation type.
This produces a clear ground-truth access decision for each role-query pair.

Based on the access decision, the expected output field is constructed in a role-aware manner. 
If the reference query is authorized under the role's policy, the instance is labeled as \textsf{allowed} and retains the original gold SQL as the expected output. Otherwise, the instance is labeled as \textsf{denied}, and the expected output is replaced with a standardized denial response.

The resulting RBAC dataset retains a consistent schema across benchmark adaptations, including (i) database identifier, (ii) schema instructions, (iii) role designation, (iv) serialized access policy (authorized tables, columns, and operations), (v) the user's natural language question, (vi) expected output (gold SQL or denial response), and (vii) query difficulty. The distribution of difficulty and denial rate is illustrated in Figure~\ref{fig:allow-deny-tikz}. In addition, we retain auxiliary metadata, such as the extracted permission requirements from $Y^\star$ and the matched policy outcomes, to support downstream analysis and verification.
With this structured synthesis framework, the system is able to transform traditional text-to-SQL resources into RBAC-aware datasets while maintaining their semantic integrity.

\vspace{1mm}
\noindent\textbf{Role and policy distribution.}
\label{sec:role-distribution}
Table~\ref{tab:policy-distribution} summarizes the resulting role and policy distribution across the three benchmarks.
Here the number \textit{\# Roles} reports the number of globally distinct role names, while \textit{Avg per DB} measures the number of roles configured in each individual database including \texttt{DataOperators}.

Access scopes are highly non-uniform: most synthesized roles remain substantially constrained, with only a small fraction of the column space accessible.
The union of all non-DataOperator roles covers 97.3\% (Spider), 95.2\% (BIRD), and 92.4\% (LiveSQLBench) of the column space on average, ensuring broad coverage without relying on a single privileged role.
The role union including \texttt{DataOperator} covers 100\% of schema columns on every Spider and BIRD database and 99.1\% on LiveSQLBench databases, any residual uncovered columns treated as universally denied.

\begin{table}[t]
\centering
\caption{Role and policy distribution.}
\label{tab:policy-distribution}
\vspace{-2mm}
\footnotesize
\begin{tabular}{@{}l ccc@{}}
\toprule
\textbf{Statistic} & \textbf{Spider} & \textbf{BIRD} & \textbf{LiveSQLBench} \\
\midrule
\# Distinct semantic role names 
  & 73
  & 45
  & 90 \\
\# Configured roles per DB, range
  & 5--8
  & 5--8
  & 6--9\\
\# Configured roles per DB, avg
  & 6.5
  & 6.7
  & 7.4 \\
Entries (\texttt{DataOperator}\%)
  & 6{,}926 (42.1)
  & 10{,}175 (43.8)
  & 4{,}401 (40.4) \\
Allow / Deny (\%)
  & 53\,/\,47
  & 35\,/\,65
  & 24\,/\,76 \\
\midrule
Col.\ cov.\ w/o \texttt{DataOperator}\textsuperscript{$\ddagger$}
  & 0/23/46/30
  & 14/38/31/17
  & 67/29/4/0 \\
Avg union cov.\ w/o \texttt{DataOperator}
  & 97.3\%
  & 95.2\%
  & 92.4\% \\
\bottomrule
\multicolumn{4}{@{}l@{}}{\scriptsize\textsuperscript{$\ddagger$}\% of role--DB pairs in bins $<$25/25--50/50--75/$\geq$75 (excl.\ \texttt{DataOperator}); see Table~\ref{tab:do-ablation} for ablation.}
\end{tabular}
\vspace{-3mm}
\end{table}

\section{Metrics and Evaluation Protocol}\label{sec:benchmarking_tools}

As mentioned in Section~\ref{sec:introduction}, a key challenge in assessing the RBAC capabilities for text-to-SQL systems is the lack of standardized metrics and evaluation protocols that jointly account for access control and SQL correctness. 
In this section, we present our evaluation framework tailored to the RBAC setting.

\subsection{Metric Design}
\label{sec:metrics}
Evaluating text-to-SQL systems in an RBAC context requires moving beyond traditional accuracy metrics toward a multi-faceted framework that measures both policy compliance (safety) and SQL code correctness (utility). 
In contrast to conventional benchmarks that mainly rely on the EX metric (explained below), we design a comprehensive suite consisting of the specialized \textit{access control (AC)} metrics, the fine-grained \textit{six-category outcome space}, and the holistic \textit{Safe-EX} metric, elaborated shortly.

\vspace{1mm}
\noindent\textbf{Limitations of execution accuracy.} 
Existing text-to-SQL benchmarks adopt the \textit{execution accuracy} (\textit{EX}) metric, which deems an output correct (referred to as \textit{execution-correct}) if its execution result matches that of the reference (gold) SQL~\cite{yu_spider_2019, li_can_2023}.
EX correctly handles semantically equivalent but syntactically different SQL queries. For example, for the question ``Show all department names,'' with gold SQL 
\lstinline[style=mystyle,language=SQL]!SELECT name FROM department!,  
the output
\lstinline[style=mystyle,language=SQL]!SELECT T1.name FROM department AS T1! 
is execution-correct, even though it does not exactly match the gold SQL textually.

However, as mentioned in Section~\ref{sec:introduction}, 
EX is insufficient in an RBAC context, since it does not cover the outcome that a query should be denied (see Section~\ref{sec:pre:text-to-sql}). 
In fact, EX can be misleading since it measures a model's coding capability but not its compliance, leading to \textit{RBAC-rejected successes} in which a
system produces syntactically-correct yet unauthorized queries, e.g., the SQL in Example~\ref{example:rbac-sql}.

\vspace{1mm}
\noindent
\textbf{Six-category outcome space.} 
To disentangle access control from SQL generation quality, we check both policy compliance and execution correctness of every generated SQL query. 
Each query falls into exactly one of the following categories:
\begin{enumerate}[leftmargin=*]
    \item \textit{\underline{Correct (C).}} 
The ground-truth decision is \textsf{allow}, and the generated SQL is policy-compliant and execution-correct. This is an ideal outcome.
    \item \textit{\underline{Wrong (W).}} 
The ground-truth decision is \textsf{allow}, and the generated SQL is policy-compliant but execution-incorrect.
    \item \textit{\underline{Proper refusal (PR).}} 
    The query was correctly denied. This is the other ideal outcome besides \textit{Correct}.
    \item \textit{\underline{Violation correct (VC).}}
The generated SQL is execution-correct, but the ground-truth decision is \textsf{deny} or the SQL violates the role policy (RBAC-rejected success).
    \item \textit{\underline{Violation wrong (VW).}} 
The generated SQL is execution-incorrect, and the ground-truth decision is \textsf{deny} or the SQL violates the role policy.
    \item \textit{\underline{Over-refusal (OR).}} 
    The query was denied despite being authorized, reflecting a security misjudgment and utility loss.
    
\end{enumerate}
This outcome space enables fine-grained diagnosis of failure modes.
For example, a high VC rate indicates that a system understands the schema but ignores access constraints, whereas a high VW rate often correlates with schema misunderstanding under restricted exposure (see Section~\ref{subsec:schema_expose}).

\vspace{1mm}
\noindent
\textbf{Access control (AC) metrics.}
Based on this categorization, we define access-control-aware (AC) metrics that explicitly evaluate authorization behavior under RBAC. 
Let $N$ denote the total number of role-query instances in the evaluation set. We define the following safety metrics:

\begin{itemize}[leftmargin=*]
    \item \underline{\textit{Violation rate}} $\downarrow$ is $(|\mathrm{VC}|+|\mathrm{VW}|)/N$, the fraction of all instances classified as VC or VW.
    \item \underline{\textit{Over-refusal rate}} $\downarrow$ is $|\mathrm{OR}|/N$, the fraction of all instances for which the ground-truth decision is \textsf{allow} but the system refuses the query. It captures utility loss from denying legitimate access, but does not by itself create a security concern.
    \item \underline{\textit{AC-F1 score}} $\uparrow$ evaluates access-control behavior using the six outcome categories. With $T=|\mathrm{C}|+|\mathrm{W}|$, $V=|\mathrm{VC}|+|\mathrm{VW}|$, and $O=|\mathrm{OR}|$, AC-F1 is $2T/(2T+V+O)$, the harmonic mean of the resulting precision and recall. C/W receive credit for compliant answers to allowed requests; VC/VW and OR penalize violating answers and over-refusals, respectively. SQL utility metrics distinguish C from W.
\end{itemize}
These metrics are orthogonal to SQL correctness and focus exclusively on whether the system's behavior aligns with RBAC policy.

\vspace{1mm}
\noindent
\textbf{Overall safety-utility metric.}
While the above metrics provide fine-grained insights, practitioners often require a single, holistic metric that summarizes the practical utility of a system.
To this end, we define \emph{Safe Execution Accuracy (Safe-EX)} as \ignore{the fraction of instances that are both RBAC-compliant and execution-correct: }the fraction of ground-truth allowed instances for which the system returns SQL that is both execution-correct and RBAC-compliant:
\[\text{Safe-EX}=\frac{C}{N_+},\]
where $N_+$ denotes the number of instances whose ground-truth access decision is \textsf{allow}, and $C$ is the number of correctly allowed instances with execution-correct SQL.
This metric directly answers an important question for a practitioner: ``How often does this system return safe, execution-correct SQL to the user?''. 
It captures the safety versus utility trade-off in a single number, where a high score requires both high SQL accuracy and strict policy adherence.
Symmetrically, we define \emph{Safe-Deny} $= \mathrm{PR}/N^{-}$ as the fraction of denied queries that are correctly refused, where $N^{-}$ denotes the number of \textsf{deny} instances. This metric complements Safe-EX by quantifying how reliably a system blocks unauthorized access. We report Safe-Deny alongside Safe-EX in the SFT study (Table~\ref{tab:rbac_spider_sft}).

\subsection{Evaluation Protocol}\label{sec:metrics_and_tools:evaluation_tools}

\noindent\textbf{Multi-pass evaluation.} 
Instead of evaluating all queries and roles, we adopt a stratified sampling strategy where each natural language query is paired with exactly one randomly selected role and its policy per evaluation pass. 
We prioritize this approach to preserve the task difficulty distribution of the original benchmarks. 
Since complex databases (often associated with ``Hard'' or ``Extra Hard'' queries) necessitate a larger number of synthesized roles to cover their schemas, evaluating all (query, role) pairs would disproportionately over-represent these difficult instances, thereby skewing the overall metrics.
To mitigate the variance introduced by random role assignment, we repeat the evaluation process with $k=5$ distinct random seeds. 
Meanwhile, we report all averaged metrics 
with standard deviation, ensuring that our results reflect robust system performance rather than sampling artifacts.

\vspace{1mm}
\noindent\textbf{Execution and pre-filtering.}
To improve efficiency, we apply a lightweight heuristic filter to exclude obvious refusal responses or non-SQL text prior to execution.
For the remaining queries, we follow the ephemeral database mechanism to ensure execution validity, particularly for LiveSQLBench which involves data modification (CRUD) operations. 
Each generated SQL is executed in an isolated transaction or temporary instance that is rolled back post-verification. 
This design isolates state changes, ensuring that the execution result of one query does not interfere with the correctness of subsequent evaluations.

\section{Experimental Evaluation}\label{sec:experiments}

This section reports experimental results for text-to-SQL under RBAC.
We run each model in two settings: standard text-to-SQL and RBAC-conditioned generation using our augmented benchmarks. 
All models share the same zero-shot prompt template and evaluation harness; the RBAC setting only adds role/policy inputs.
We report utility and safety metrics as defined in Section~\ref{sec:metrics}, and Section~\ref{sec:empirical-analysis} analyzes failure modes in detail.

\subsection{Setup}
\label{sec:exp-settings}

As shown in Figure~\ref{fig:overview}, to reduce bias from model-specific prompt strategies and example styles, we adopt a consistent task instruction. 
For selected models, we conduct two sets of experiments: (i) standard text-to-SQL evaluation without access control, and (ii) access control-aware evaluation using our RBAC-augmented benchmarks. 
All queries are executed using an aligned backend including SQLite and PostgreSQL to ensure consistent and lightweight execution across datasets.

\vspace{1mm}
\noindent
\textbf{Models.} 
We selected a comprehensive and representative set of models spanning different deployment settings, training paradigms, and model scales, including both general-purpose language models and specialized text-to-SQL models. The evaluated models include: 
\begin{itemize}[leftmargin=*]
\item Commercial models, including Google Gemini-2.5-Flash \cite{comanici2025gemini}, Anthropic Claude-Sonnet-4.5-20250929~\cite{Claude-Sonnet}, and OpenAI GPT-4o-mini, GPT-5-mini and GPT-5~\cite{openai-models}.
\item Open-weight general-purpose models, including DeepSeek V3.2 \cite{liu2024deepseek} (both Reasoning and Coder variants), Gemma-3 4B and 27B \cite{team2025gemma}, Qwen-2.5 14B-Instruct~\cite{qwen2024qwen25technicalreport} and Coder-7B-Instruct~\cite{hui2024qwen2}.
\item Specialized text-to-SQL models, including  
Llama3-SQLCoder-8B \cite{llama-3-sqlcoder-8b} 
and Snowflake-Arctic-R1-7B \cite{yao2025arctic}.

\end{itemize}

All models are evaluated without supervised fine-tuning by default and are provided with the same zero-shot template, 
a fixed temperature setting of 0 when supported by the provider, and a unified execution pipeline with dataset-specific backends (SQLite for Spider and BIRD, and PostgreSQL for LiveSQLBench). 
Model names follow the providers' public identifiers.
All artifacts are available in our public GitHub repository~\cite{yang2026rbactext2sql}.

\subsection{Overall Performance}\label{sec:exp:performance}

\begingroup
\setlength{\tabcolsep}{6pt}
\newcolumntype{V}{!{\vrule width .5pt}}

\let\oldopenrow\openrow
\let\oldclosedrow\closedrow
\renewcommand{\openrow}{\oldopenrow\cellcolor{white}}
\renewcommand{\closedrow}{\oldclosedrow\cellcolor{white}}

\begin{table*}[htbp]
\caption{Overall Performance on Spider, BIRD, and LiveSQLBench considering RBAC.
\textbf{Bold} marks the best and \underline{underline} the second-best value within each dataset block.
\legendbox{OpenBG}{Open-weight LLM} rows and \legendbox{ClosedBG}{commercial LLM} rows have different color shades.
}
\vspace{-2mm}

\centering
\small

\begin{tabulary}{\textwidth}{L V J C C V C C C V C C}
\toprule
\multicolumn{1}{l}{\multirow{2}{*}{\textbf{Dataset}}} &
\multicolumn{1}{l}{\multirow{2}{*}{\textbf{\ \ Model}}}   &
\multicolumn{1}{l}{\textbf{EX$\uparrow$}} &
\multicolumn{1}{l}{\textbf{Safe-EX$\uparrow$}} &
\multicolumn{1}{l}{\textbf{Violation$\downarrow$}} &
\multicolumn{1}{l}{\textbf{OverRefusal$\downarrow$}} &
\multicolumn{1}{l}{\textbf{AC-F1$\uparrow$}} &
\multicolumn{2}{c}{\makecell{\textbf{Avg. Cost/Task} \\(USD $\times 10^{-3}$)}} \\
\cmidrule(lr){3-9}
\multicolumn{2}{c}{} &
\textbf{w/o role} &
\textbf{w/ role} &
 &
\textbf{w/ role} &
 &
\textbf{w/o role} &
\textbf{w/ role} \\
\midrule

\openrow
& Snowflake-R1-7b             & 78.14 & 76.33 $\pm$ 1.60 & 48.68 $\pm$ 1.00 & 0.08 $\pm$ 0.07 & 67.31 $\pm$ 0.93 & -- & -- \\
\openrow
& Llama3-SQLCoder-8b        & 58.70 & 60.52 $\pm$ 1.07 & 50.41 $\pm$ 0.98 & 0.00 $\pm$ 0.00 & 66.30 $\pm$ 0.87 & -- & -- \\
\openrow
& Gemma3-4b                 & 68.76 & 68.34 $\pm$ 1.25 & 50.19 $\pm$ 1.24 & 0.00 $\pm$ 0.00 & 66.46 $\pm$ 1.09 & 0.016 & 0.017 \\
\openrow
& Gemma3-27b                & 80.08 & 77.15 $\pm$ 1.63 & 34.14 $\pm$ 1.09 & 0.93 $\pm$ 0.10 & 74.10 $\pm$ 1.05 & 0.036 & 0.037 \\
\openrow
& Qwen2.5-Coder-7b-Inst. & 74.66 & 72.51 $\pm$ 0.28 & 49.65 $\pm$ 1.14 & 0.00 $\pm$ 0.00 & 66.91 $\pm$ 1.03 & 0.057 & 0.059 \\
\openrow
& Qwen2.5-14b-Instruct      & 70.21 & 68.74 $\pm$ 0.81 & 36.94 $\pm$ 1.72 & 0.54 $\pm$ 0.20 & 72.84 $\pm$ 1.21 & 0.060 & 0.062 \\
\openrow
& Deepseek/V3.2-Reasoning   & 78.82 & 75.62 $\pm$ 1.16 & 2.11 $\pm$ 0.30 & 2.36 $\pm$ 0.33 & \textbf{95.80 $\pm$ 0.44} & 0.112 & 0.115 \\
\openrow
& Deepseek/V3.2-Coder       & 76.98 & 71.17 $\pm$ 1.08 & 33.60 $\pm$ 1.06 & 0.58 $\pm$ 0.18 & 74.81 $\pm$ 0.93 & 0.112 & 0.115 \\
\closedrow
& Claude-Sonnet-4.5         & \textbf{85.40} & \textbf{80.52 $\pm$ 0.90} & 4.89 $\pm$ 0.18 & 3.01 $\pm$ 0.42 & 92.68 $\pm$ 0.50 & 1.420 & 1.448 \\
\closedrow
& Gemini-2.5-Flash          & \underline{84.24} & \underline{80.33 $\pm$ 1.52} & 10.87 $\pm$ 0.22 & 2.09 $\pm$ 0.25 & 88.66 $\pm$ 0.39 & 0.163 & 0.166 \\
\closedrow
& OpenAI/GPT-4o-mini        & 74.76 & 60.11 $\pm$ 1.08 & 17.00 $\pm$ 1.35 & 10.37 $\pm$ 0.53 & 75.31 $\pm$ 1.72 & 0.068 & 0.069 \\
\closedrow
& OpenAI/GPT-5-mini         & 75.15 & 71.25 $\pm$ 0.63 & 3.60 $\pm$ 0.50 & 1.72 $\pm$ 0.32 & 95.10 $\pm$ 0.34 & 0.134 & 0.136 \\
\closedrow
\multirow{-13}{*}{\makecell[l]{Spider}} & OpenAI/GPT-5              & 74.37 & 67.62 $\pm$ 1.36 & 2.61 $\pm$ 0.37 & 1.97 $\pm$ 0.31 & \underline{95.74 $\pm$ 0.33} & 0.670 & 0.682 \\
\midrule

\openrow
& Snowflake-R1-7b             & 41.88 & 61.23 $\pm$ 1.14 & 65.10 $\pm$ 0.86 & 0.07 $\pm$ 0.07 & 51.45 $\pm$ 1.01 & -- & -- \\
\openrow
& Llama3-SQLCoder-8b        & 21.92 & 38.21 $\pm$ 1.52 & 66.19 $\pm$ 1.04 & 0.00 $\pm$ 0.00 & 50.49 $\pm$ 1.18 & -- & -- \\
\openrow
& Gemma3-4b                 & 22.18 & 36.97 $\pm$ 1.12 & 66.42 $\pm$ 0.87 & 0.00 $\pm$ 0.00 & 50.22 $\pm$ 0.97 & 0.043 & 0.044 \\
\openrow
& Gemma3-27b                & 40.70 & 49.92 $\pm$ 1.52 & 33.42 $\pm$ 0.72 & 1.66 $\pm$ 0.25 & 65.67 $\pm$ 1.02 & 0.097 & 0.098 \\
\openrow
& Qwen2.5-Coder-7b-Inst. & 31.77 & 44.99 $\pm$ 1.94 & 65.94 $\pm$ 0.98 & 0.00 $\pm$ 0.00 & 50.75 $\pm$ 1.08 & 0.152 & 0.154 \\
\openrow
& Qwen2.5-14b-Instruct      & 28.70 & 45.14 $\pm$ 1.07 & 56.19 $\pm$ 0.92 & 0.13 $\pm$ 0.07 & 54.64 $\pm$ 0.94 & 0.157 & 0.158 \\
\openrow
& Deepseek/V3.2-Reasoning   & 39.79 & 56.69 $\pm$ 1.25 & 8.82 $\pm$ 0.46 & 1.37 $\pm$ 0.16 & \underline{87.10 $\pm$ 0.55} & 0.300 & 0.302 \\
\openrow
& Deepseek/V3.2-Coder       & 38.29 & 55.23 $\pm$ 1.51 & 49.90 $\pm$ 1.12 & 0.13 $\pm$ 0.08 & 57.89 $\pm$ 1.16 & 0.300 & 0.302 \\
\closedrow
& Claude-Sonnet-4.5         & \textbf{53.69} & \underline{64.73 $\pm$ 0.78} & 7.70 $\pm$ 0.50 & 1.81 $\pm$ 0.15 & \textbf{87.65 $\pm$ 0.92} & 3.571 & 3.599 \\
\closedrow
& Gemini-2.5-Flash          & \underline{51.66} & \textbf{66.61 $\pm$ 0.94} & 21.23 $\pm$ 1.13 & 0.91 $\pm$ 0.14 & 75.59 $\pm$ 1.33 & 0.391 & 0.394 \\
\closedrow
& OpenAI/GPT-4o-mini        & 36.42 & 40.61 $\pm$ 0.98 & 29.66 $\pm$ 0.81 & 5.05 $\pm$ 0.49 & 63.18 $\pm$ 0.69 & 0.173 & 0.175 \\
\closedrow
& OpenAI/GPT-5-mini         & 41.36 & 56.43 $\pm$ 0.91 & 12.68 $\pm$ 0.50 & 0.65 $\pm$ 0.14 & 84.03 $\pm$ 0.46 & 0.323 & 0.326 \\
\closedrow
\multirow{-13}{*}{\makecell[l]{BIRD}} & OpenAI/GPT-5              & 43.05 & 59.25 $\pm$ 1.56 & 10.49 $\pm$ 0.69 & 0.68 $\pm$ 0.11 & 86.32 $\pm$ 0.73 & 1.617 & 1.629 \\
\midrule

\openrow
& Snowflake-R1-7b             & 5.91 & 10.86 $\pm$ 2.98 & 82.03 $\pm$ 1.37 & 0.27 $\pm$ 0.08 & 29.33 $\pm$ 1.91 & -- & -- \\
\openrow
& Gemma3-4b                 & 3.38 & 5.83 $\pm$ 1.68 & 72.87 $\pm$ 0.55 & 2.57 $\pm$ 0.20 & 28.89 $\pm$ 1.21 & 0.393 & 0.394 \\
\openrow
& Gemma3-27b                & 8.45 & 10.34 $\pm$ 1.56 & 60.44 $\pm$ 1.05 & 1.93 $\pm$ 0.36 & 34.12 $\pm$ 1.00 & 0.884 & 0.884 \\
\openrow
& Qwen2.5-Coder-7b-Inst. & 5.07 & 9.05 $\pm$ 1.21 & 73.17 $\pm$ 1.47 & 0.84 $\pm$ 0.24 & 31.86 $\pm$ 2.23 & 1.377 & 1.378 \\
\openrow
& Qwen2.5-14b-Instruct      & 7.26 & 10.51 $\pm$ 1.70 & 59.43 $\pm$ 1.50 & 1.11 $\pm$ 0.17 & 37.46 $\pm$ 1.78 & 1.386 & 1.388 \\
\openrow
& Deepseek/V3.2-Reasoning   & \underline{23.14} & 22.01 $\pm$ 3.26 & 26.82 $\pm$ 1.26 & 6.11 $\pm$ 0.64 & 48.80 $\pm$ 2.20 & 2.743 & 2.746 \\
\openrow
& Deepseek/V3.2-Coder       & 18.92 & 19.49 $\pm$ 1.45 & 45.17 $\pm$ 1.33 & 2.80 $\pm$ 0.70 & 38.47 $\pm$ 2.04 & 2.743 & 2.746 \\
\closedrow
& Claude-Sonnet-4.5         & 21.79 \hspace{1em} & 18.41 $\pm$ 2.99 \hspace{1em} & 13.28 $\pm$ 1.86 \hspace{1em} & 8.58 $\pm$ 0.56 \hspace{2em} & 55.75 $\pm$ 3.57 & 30.136 & 30.165 \\
\closedrow
& Gemini-2.5-Flash          & 21.28 & 16.24 $\pm$ 3.36 & 35.31 $\pm$ 2.04 & 3.78 $\pm$ 0.94 & 44.89 $\pm$ 2.54 & 3.085 & 3.088 \\
\closedrow
& OpenAI/GPT-4o-mini        & 13.01 & 8.09 $\pm$ 2.25 & 25.37 $\pm$ 0.86 & 12.23 $\pm$ 1.55 & 30.46 $\pm$ 2.54 & 1.496 & 1.498 \\
\closedrow
& OpenAI/GPT-5-mini         & 18.58 & \underline{25.57 $\pm$ 0.91} & 19.70 $\pm$ 1.61 & 4.32 $\pm$ 0.31 & \underline{58.49 $\pm$ 2.51} & 2.565 & 2.567 \\
\closedrow
\multirow{-12}{*}{\makecell[l]{LiveSQL-\\Bench}} & OpenAI/GPT-5              & \textbf{27.20} & \textbf{28.66 $\pm$ 2.39} & 17.84 $\pm$ 1.78 & 4.53 $\pm$ 0.64 & \textbf{61.15 $\pm$ 2.37} & 12.823 & 12.835 \\

\bottomrule
\end{tabulary}
\label{tab:overall}
\end{table*}
\endgroup

We evaluate each model along two dimensions: (i) utility, measured by EX in the conventional setting and Safe-EX under RBAC, and (ii) safety, measured by AC metrics, including Violation rate, OverRefusal rate, and AC-F1, as defined in Section~\ref{sec:metrics}. Table~\ref{tab:overall} reports the overall results. 
Since EX is measured on the original unrestricted benchmark, whereas Safe-EX is measured only on ground-truth allowed instances in the RBAC-augmented benchmark, the two metrics are not directly comparable as raw percentages. 
Our EX scores closely match publicly reported single-model results on Spider, BIRD, and LiveSQLBench under zero-shot prompt settings.

\vspace{1mm}
\noindent
\textbf{Access control results.} Under RBAC, AC-F1 varies substantially across models and datasets, with violation rates consistently exceeding over-refusal rates. On Spider, the flagship commercial model GPT-5 attains strong AC-F1 results and relatively low (but still non-zero) violation rates.
In contrast, several open-weight models show significantly higher violation rates despite their similar or higher EX scores compared to GPT-5. On BIRD and LiveSQLBench, all evaluated models show degraded AC-F1 and elevated violation rates. 
For instance, Snowflake-R1-7b, a strong model in terms of EX scores, records a {65.10\%} violation rate on BIRD, while multiple models on LiveSQLBench retain double-digit violation rate alongside low AC-F1. 
These results indicate that enforcing RBAC remains challenging even for strong text-to-SQL models.

\vspace{1mm}
\noindent
\textbf{Utility results.}
Safe-EX consistently declines from Spider to BIRD and further on LiveSQLBench, reflecting increasing schema complexity and stricter access constraints.
Moreover, although EX and Safe-EX use different denominators, with EX measured on the original unrestricted benchmark and Safe-EX on RBAC-allowed augmented instances, the results still show that high unrestricted SQL accuracy does not necessarily translate to robust performance under access control.
Several models with competitive EX achieve only modest Safe-EX once RBAC is enforced.

\vspace{1mm}
\noindent\textbf{Utility-safety trade-offs.} 
We observe a systematic tradeoff between SQL utility and access-control safety when moving from the base model to their reasoning-oriented variants in Table~\ref{tab:overall}. Using Safe-EX and AC-F1 as utility and safety metrics, respectively, Snowflake-R1-7B consistently improves utility over its base model Qwen2.5-Coder-7B-Instruct on all three benchmarks, yet the former's safety is no better than the latter's. This pattern is most pronounced on LiveSQLBench, where Safe-EX increases but AC-F1 declines, indicating that reasoning-oriented post-training prioritizes executable SQL generation under constraints, but weakens refusal alignment at decision time. This confirms that better SQL execution under RBAC does not automatically confer better policy compliance, and motivates joint objectives or decoding controls that couple reasoning with calibrated denial behaviors.

\vspace{1mm}
\noindent\textbf{Commercial vs. open-weight LLMs.} 
A clear separation emerges between commercial and open-weight models. Commercial models generally achieve higher AC-F1 scores and lower violation rates than open-weight baselines, although violations remain nontrivial on harder datasets such as BIRD and LiveSQLBench. In contrast, many open-weight models, including text-to-SQL-specialized models with strong EX performance on standard leaderboards, exhibit substantially higher violation rates. 
This gap may partly reflect differences in pre-training and post-training objectives.
Commercial models are typically trained with extensive instruction tuning and safety alignment, which may partially transfer to RBAC compliance. By contrast, open-weight text-to-SQL models are often optimized primarily for SQL accuracy, which may leave them less robust to access-control constraints.

\vspace{1mm}
\noindent\textbf{Task complexity and access control.} 
We further analyze the effect of task difficulty on RBAC compliance by grouping queries according to the difficulty annotations of the source benchmarks. 
Table \ref{tab:bird-singlecol-main} reports representative results on BIRD. Due to space constraints, full results covering additional models and all three benchmarks are provided in Appendix~\ref{appendix:per-difficulty}. 
Across all benchmarks and models, we observe a clear and consistent inverse relationship between task complexity and RBAC compliance: as queries progress from easier to harder categories, safety performance (i.e., AC-F1) consistently degrades. 
Notably, this degradation is driven primarily by increased RBAC violations rather than over-refusals, indicating that under higher cognitive load, models tend to prioritize SQL generation utility over access control constraints.

\begin{figure}[t]
    \centering
    \includegraphics[width=0.95\linewidth, trim=0.2cm 0.75cm 0cm 0.2cm, clip]{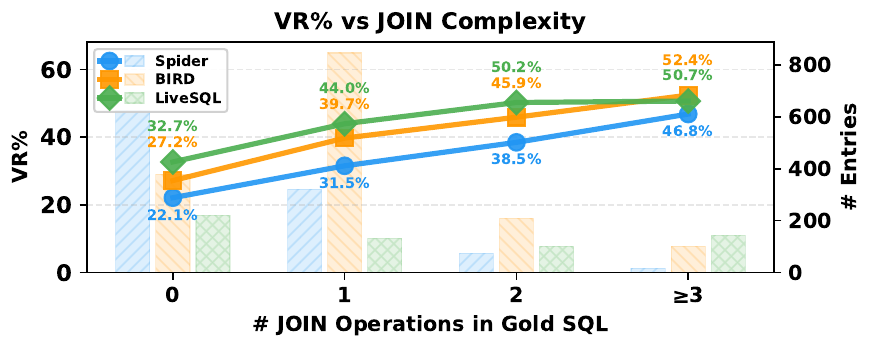}
    \vspace{-4mm}
    \caption{Violation rate vs.\ \# \texttt{JOIN} operations in the gold SQL.}
    \label{fig:multi-table-revised}
    \vspace{-3mm}
\end{figure}

\vspace{1mm}
\noindent\textbf{Multi-table join complexity.}
We further stratify violation behavior by the number of \texttt{JOIN} operations in gold SQL.
Figure~\ref{fig:multi-table-revised} shows the average results for all tested models. 0-\texttt{JOIN} bin contains queries without any \texttt{JOIN} clause: single-table \texttt{SELECT}s on Spider/BIRD, and additionally CRUD queries on LiveSQLBench.
The violation rate increases with the number of \texttt{JOIN}s, consistent with the expectation that cross-table reasoning makes policy compliance harder.

\vspace{1mm}
\noindent\textbf{Effect of scoped administrators.}
Table~\ref{tab:do-ablation} compares model performance on Spider with and without \texttt{DataOperator} roles.
Removing \texttt{DataOperator} lowers the violation rate, since it removes many non-trivial deny-case instances, but model rankings remain largely stable. 
This indicates that scoped administrators make the benchmark harder by adding genuine access-control decisions, without distorting relative model comparisons. 

\vspace{1mm}
\noindent\textbf{Granularity and adaptability.}
Our benchmark currently covers standard \emph{column-level} and \emph{operation-level} RBAC, which are granularities supported by many relational DBMSs.
Finer-grained \emph{row/cell-level} policies are not included in the main release due to their substantially higher synthesis and human validation costs, and are left as future work. 
As a limited feasibility check, we further conducted a row-level study to verify that our benchmarking framework can be extended to row-level predicates; details and results are deferred to Appendix~\ref{appendix:row-level}.

\begin{table}[t]
\centering
\caption{Model performance with/without \texttt{DataOperator}.}
\vspace{-2mm}
\label{tab:do-ablation}
\footnotesize
\setlength{\tabcolsep}{2.5pt}
\begin{tabular}{@{}l cc cc cc cc@{}}
\toprule
\multirow{2}{*}{\textbf{Model}}
& \multicolumn{2}{c}{\textbf{Safe-EX}$\uparrow$}
& \multicolumn{2}{c}{\textbf{Viol-Rate(\%)}$\downarrow$}
& \multicolumn{2}{c}{\textbf{OR-Rate(\%)}$\downarrow$}
& \multicolumn{2}{c}{\textbf{AC-F1}$\uparrow$} \\
\cmidrule(lr){2-3}\cmidrule(lr){4-5}\cmidrule(lr){6-7}\cmidrule(lr){8-9}
& w/ & w/o & w/ & w/o & w/ & w/o & w/ & w/o \\
\midrule

DeepSeek-Coder    & 71.2 & 71.2 & 33.6 & 23.6 & 0.6 & 0.8 & 74.8 & 81.7 \\
Claude-Sonnet-4.5 & 80.5 & 82.6 & 4.9  & 4.6  & 3.0 & 2.5 & 92.7 & 93.7 \\
GPT-4o-mini       & 60.1 & 61.6 & 17.0 & 10.9 & 10.4 & 10.7 & 75.3 & 80.8 \\
GPT-5             & 67.6 & 69.2 & 2.6  & 2.5  & 2.0 & 1.8 & 95.7 & 96.2 \\

\bottomrule
\end{tabular}
\vspace{-3mm}
\end{table}

\begin{table}[htbp]
\caption{RBAC-BIRD difficulty tier performance.}
\vspace{-2mm}
\centering
\footnotesize
\setlength{\tabcolsep}{3.9pt}
\begin{adjustbox}{width=1.0\columnwidth,center}
\begin{tabulary}{\columnwidth}{L C C C C C}
\toprule
\textbf{Model} & \textbf{Difficulty}  & \textbf{Safe-EX$\uparrow$} & \textbf{Viol. (\%)$\downarrow$} & \textbf{OR (\%)$\downarrow$} & \textbf{AC-F1$\uparrow$} \\
\midrule

\multirow{3}{*}{\makecell[l]{Snowflake- \\ R1-7b}}
  & Simple    & 68.03$\pm$0.87 & 59.95$\pm$1.46 & 0.05$\pm$0.06 & 56.85$\pm$1.56 \\
  & Moderate  & 49.75$\pm$1.73 & 66.77$\pm$1.02 & 0.14$\pm$0.18 & 49.65$\pm$1.30 \\
  & Challenge & 47.01$\pm$6.00 & 81.04$\pm$1.21 & 0.00$\pm$0.00 & 31.65$\pm$1.93 \\
\cmidrule(lr){2-6}

\multirow{3}{*}{\makecell[l]{Llama3- \\ SQLCoder-8b}}
  & Simple    & 47.61$\pm$2.03 & 61.23$\pm$1.59 & 0.00$\pm$0.00 & 55.79$\pm$1.69 \\
  & Moderate  & 21.39$\pm$3.81 & 67.95$\pm$1.04 & 0.00$\pm$0.00 & 48.54$\pm$1.20 \\
  & Challenge & 21.39$\pm$3.00 & 81.21$\pm$1.05 & 0.00$\pm$0.00 & 31.62$\pm$1.48 \\
\cmidrule(lr){2-6}

\multirow{3}{*}{\makecell[l]{Gemma3-27b}}
  & Simple    & 58.64$\pm$1.04 & 30.90$\pm$1.76 & 1.91$\pm$0.45 & 70.06$\pm$2.01 \\
  & Moderate  & 34.89$\pm$4.27 & 36.79$\pm$1.84 & 1.76$\pm$0.26 & 62.29$\pm$1.50 \\
  & Challenge & 32.51$\pm$5.59 & 36.36$\pm$1.32 & 0.52$\pm$0.51 & 50.45$\pm$1.64 \\
\cmidrule(lr){2-6}

\multirow{3}{*}{\makecell[l]{Qwen2.5- \\ Coder-7b}}
  & Simple    & 54.40$\pm$1.38 & 60.82$\pm$1.60 & 0.00$\pm$0.00 & 56.24$\pm$1.65 \\
  & Moderate  & 29.49$\pm$3.94 & 67.90$\pm$1.12 & 0.00$\pm$0.00 & 48.52$\pm$1.29 \\
  & Challenge & 24.08$\pm$4.13 & 81.21$\pm$1.52 & 0.00$\pm$0.00 & 31.51$\pm$2.05 \\
\cmidrule(lr){2-6}

\multirow{3}{*}{\makecell[l]{Qwen2.5-14b}}
  & Simple    & 53.14$\pm$1.07 & 50.71$\pm$1.23 & 0.14$\pm$0.09 & 60.47$\pm$1.48 \\
  & Moderate  & 32.10$\pm$3.06 & 59.19$\pm$1.89 & 0.14$\pm$0.11 & 52.28$\pm$1.59 \\
  & Challenge & 26.65$\pm$2.32 & 70.82$\pm$2.50 & 0.09$\pm$0.17 & 33.78$\pm$2.38 \\
\cmidrule(lr){2-6}

\multirow{3}{*}{\makecell[l]{Deepseek- \\ V3.2-Coder}}
  & Simple    & 61.82$\pm$1.19 & 44.17$\pm$1.33 & 0.19$\pm$0.16 & 64.09$\pm$1.61 \\
  & Moderate  & 47.15$\pm$3.68 & 50.48$\pm$2.11 & 0.09$\pm$0.11 & 56.55$\pm$1.76 \\
  & Challenge & 31.26$\pm$6.64 & 70.13$\pm$1.88 & 0.00$\pm$0.00 & 33.69$\pm$2.48 \\
\cmidrule(lr){2-6}

\multirow{3}{*}{\makecell[l]{Claude- \\ Sonnet-4.5}}
  & Simple    & 70.14$\pm$0.43 & 7.45$\pm$0.45 & 1.91$\pm$0.23 & 89.22$\pm$0.58 \\
  & Moderate  & 57.01$\pm$3.60 & 7.99$\pm$1.04 & 1.81$\pm$0.57 & 86.79$\pm$1.40 \\
  & Challenge & 48.72$\pm$7.58 & 8.05$\pm$1.05 & 1.47$\pm$0.70 & 79.16$\pm$3.68 \\
\cmidrule(lr){2-6}

\multirow{3}{*}{\makecell[l]{Gemini- \\ 2.5-Flash}}
  & Simple    & 72.85$\pm$0.75 & 19.00$\pm$1.24 & 0.95$\pm$0.15 & 79.79$\pm$1.10 \\
  & Moderate  & 57.34$\pm$2.29 & 23.21$\pm$2.69 & 0.95$\pm$0.36 & 73.02$\pm$2.62 \\
  & Challenge & 49.39$\pm$5.73 & 25.72$\pm$2.33 & 0.69$\pm$0.44 & 58.27$\pm$3.14 \\
\cmidrule(lr){2-6}

\multirow{3}{*}{\makecell[l]{GPT-5}}
& Simple    & 68.60$\pm$1.62 & 9.36$\pm$0.90 & 0.86$\pm$0.09 & 88.73$\pm$0.65 \\
& Moderate  & 44.39$\pm$5.50 & 11.20$\pm$1.20 & 0.41$\pm$0.26 & 85.39$\pm$1.51 \\
& Challenge & 36.57$\pm$7.66 & 13.33$\pm$1.04 & 0.52$\pm$0.33 & 73.55$\pm$2.62 \\
\bottomrule
\end{tabulary}
\end{adjustbox}
\label{tab:bird-singlecol-main}
\vspace{-3mm}
\end{table}

\section{Empirical Analysis of RBAC Compliance}\label{sec:empirical-analysis}

This section presents a detailed empirical analysis of text-to-SQL behavior under RBAC constraints.
Beyond aggregate performance metrics, the analysis focuses on systematic factors that influence access-control compliance during SQL generation.
Specifically, we analyze the impact of schema exposure, characterize representative access-control failures in model reasoning, and evaluate the effectiveness and limitations of two common heuristic remedies: \textit{supervised fine-tuning} and \textit{in-context learning}.

\subsection{Impact of Schema Exposure}
\label{subsec:schema_expose}
\begin{table*}[!t]
\centering
\footnotesize

\caption{Comparison of
\emph{Full-Schema} vs \emph{Role-Schema} on BIRD RBAC Data.}
\vspace{-2mm}

\label{tab:schema_exposure}

\begin{tabulary}{\textwidth}{L L R R R R R R R R}
\toprule
\textbf{Model} & \textbf{Setting} & \textbf{$\#$ValidG} & \textbf{$\#$InvalidG} & \textbf{$\#$FailtoRefuse} & \textbf{Viol (\%)} & \textbf{VC} & \textbf{VW} & \textbf{OR (\%)} & \textbf{AC-F1} \\
\midrule

\multirow{3}{*}{\makecell[l]{Snowflake-R1-7b}}
& Full\mbox{-}Schema  & 589.4$\pm$9.6 & 6.2$\pm$0.7 & 402.2$\pm$12.3 & 65.09$\pm$0.83 & 494.4$\pm$16.1 & 503.4$\pm$14.2 & 0.07$\pm$0.07 & 51.46$\pm$0.99 \\
& Role\mbox{-}Schema  & 236.4$\pm$11.6 & 14.8$\pm$3.7 & 596.8$\pm$10.3 & 55.32$\pm$1.05 & 25.8$\pm$2.2 & 822.2$\pm$14.8 & 0.05$\pm$0.05 & 56.36$\pm$1.12 \\
& $\Delta$
& $-$353.0$\pm$15.0 & $+$8.6$\pm$3.8
& $+$194.6$\pm$16.1
& $-$9.77$\pm$1.34
& $-$468.6$\pm$16.2
& $+$318.8$\pm$20.6
& $-$0.02$\pm$0.09
& $+$4.90$\pm$1.49 \\

\cmidrule(lr){2-10}\addlinespace[2pt]

\multirow{3}{*}{\makecell[l]{Gemma-27b}}
& Full\mbox{-}Schema  & 29.0$\pm$6.5 & 2.0$\pm$0.9 & 481.2$\pm$12.2 & 33.41$\pm$0.74 & 155.4$\pm$7.9 & 356.8$\pm$10.1 & 1.66$\pm$0.25 & 65.69$\pm$1.04 \\
& Role\mbox{-}Schema  & 22.6$\pm$2.2 & 3.2$\pm$1.2 & 425.6$\pm$18.0 & 29.45$\pm$1.22 & 32.6$\pm$2.6 & 418.8$\pm$20.2 & 1.51$\pm$0.25 & 68.92$\pm$1.21 \\
& $\Delta$
& $-$6.4$\pm$6.9 & $+$1.2$\pm$1.5
& $-$55.6$\pm$21.7
& $-$3.97$\pm$1.43
& $-$122.8$\pm$8.3
& $+$62.0$\pm$22.6
& $-$0.15$\pm$0.35
& $+$3.24$\pm$1.60 \\

\cmidrule(lr){2-10}\addlinespace[2pt]

\multirow{3}{*}{\makecell[l]{Qwen2.5-14b-Instruct}}
& Full\mbox{-}Schema  & 516.2$\pm$17.2 & 8.0$\pm$2.7 & 337.2$\pm$13.9 & 56.19$\pm$0.92 & 299.8$\pm$6.4 & 561.6$\pm$12.1 & 0.13$\pm$0.07 & 54.64$\pm$0.94 \\
& Role\mbox{-}Schema  & 247.0$\pm$11.6 & 36.0$\pm$6.6 & 379.8$\pm$15.6 & 43.24$\pm$1.01 & 30.4$\pm$3.1 & 632.4$\pm$13.3 & 0.20$\pm$0.10 & 62.01$\pm$1.00 \\
& $\Delta$
& $-$269.2$\pm$20.7 & $+$28.0$\pm$7.1
& $+$42.6$\pm$20.9
& $-$12.95$\pm$1.36
& $-$269.4$\pm$7.1
& $+$70.8$\pm$18.0
& $+$0.07$\pm$0.12
& $+$7.37$\pm$1.37 \\

\cmidrule(lr){2-10}\addlinespace[2pt]

\multirow{3}{*}{\makecell[l]{Deepseek-Coder}}
& Full\mbox{-}Schema  & 494.6$\pm$15.4 & 1.4$\pm$1.5 & 269.0$\pm$10.4 & 49.90$\pm$1.12 & 329.0$\pm$17.5 & 436.0$\pm$7.3 & 0.13$\pm$0.08 & 57.89$\pm$1.17 \\
& Role\mbox{-}Schema  & 244.8$\pm$6.8 & 9.8$\pm$3.1 & 359.6$\pm$6.8 & 40.07$\pm$0.88 & 53.6$\pm$6.9 & 560.6$\pm$15.1 & 0.59$\pm$0.07 & 63.21$\pm$0.94 \\
& $\Delta$
& $-$249.8$\pm$16.8 & $+$8.4$\pm$3.4
& $+$90.6$\pm$12.4
& $-$9.84$\pm$1.43
& $-$275.4$\pm$18.8
& $+$124.6$\pm$16.8
& $+$0.46$\pm$0.11
& $+$5.32$\pm$1.50 \\

\cmidrule(lr){2-10}\addlinespace[2pt]

\multirow{3}{*}{\makecell[l]{GPT-5-mini}}
& Full\mbox{-}Schema  & 68.6$\pm$3.8 & 0.2$\pm$0.4 & 125.6$\pm$9.7 & 12.68$\pm$0.50 & 43.0$\pm$5.4 & 151.4$\pm$2.6 & 0.65$\pm$0.14 & 84.03$\pm$0.46 \\
& Role\mbox{-}Schema  & 144.0$\pm$8.1 & 0.0$\pm$0.0 & 266.4$\pm$8.0 & 26.77$\pm$0.35 & 30.4$\pm$3.6 & 380.0$\pm$5.4 & 0.33$\pm$0.22 & 72.40$\pm$0.60 \\
& $\Delta$
& $+$75.4$\pm$8.9 & $-$0.2$\pm$0.4
& $+$140.8$\pm$12.6
& $+$14.09$\pm$0.61
& $-$12.6$\pm$6.5
& $+$228.6$\pm$6.0
& $-$0.32$\pm$0.26
& $-$11.63$\pm$0.76 \\

\bottomrule
\end{tabulary}

\vspace{-2mm}
\end{table*}
This subsection examines whether RBAC compliance in text-to-SQL can be enforced by simply removing descriptions of unauthorized schema elements. 
To this end, we evaluate the impact of schema exposure under two different settings: \emph{Full-Schema}, where the model has access to the entire database schema, and \emph{Role-Schema}, where only schema elements accessible to the role are provided. Table~\ref{tab:schema_exposure} reports the results and summarizes the following violation types:
\begin{itemize}[leftmargin=*]
    \item {\textbf{Valid-Guess}}: attempting to access valid but unauthorized columns;
    \item {\textbf{Invalid-Guess}}: attempting to access non-existent columns; 
    \item {\textbf{Fail-to-Refuse}}: generating a SQL query using only authorized columns when the query should instead be refused. 
\end{itemize}

Inspection of the violation cases shows that under the Full-Schema setting, a large number of violations involve accessing known unauthorized columns.
For example, DeepSeek-Coder records an average of {494.6} such cases.
In contrast, under the Role-Schema setting, accesses to known unauthorized columns occur much less frequently, with an average of {244.8} cases, which is likely due to the absence of schema visibility for those columns. However, rather than refusing the query, models tend to generate SQL using authorized columns and, in some cases, hallucinated ones, resulting in an increase in \emph{violation wrong} cases.
A closer examination of the detailed results shows that under the Role-Schema setting, the number of \textit{violation correct} cases (see Section~\ref{sec:metrics}) drops substantially for several models.
For example, DeepSeek-Coder shows a reduction from {329.0} to {53.6} cases, and GPT-5-mini decreases from {43.0} to {30.4} cases.
At the same time, both the number of {violation wrong} cases and the overall violation rate increase for strong models such as GPT-5-mini, with the violation count rising from {151.4} to {380.0}, corresponding to an increase of {14.1} percentage points.

In sum, these results indicate that simply restricting schema visibility to role-accessible columns reduces explicit data leakage but does not effectively enforce RBAC policies, as models continue to hallucinate and rarely refuse unauthorized queries.

\subsection{Case Study: RBAC Failures in Reasoning}
\label{subsec:reason-case}

We observe a characteristic failure mode in Snowflake-R1-7b, a state-of-the-art reasoning-first model, where the final output contradicts the model's own intermediate reasoning.
Specifically, the model correctly reasons about access constraints and the requested operation, yet emits an SQL query that violates the intended RBAC decision at generation time.

\begin{figure}[t]
\raggedright
\small

\begin{mdframed}[linewidth=0.6pt,innerleftmargin=6pt,
innerrightmargin=6pt,innertopmargin=6pt,innerbottommargin=6pt]

\noindent\textbf{Model reasoning:}
To translate the given natural language question into an executable SQL query, we need to focus on the following points:

1. Identify records from the \texttt{AuditAndCompliance} table that have more than half of their findings as critical (\texttt{critFindNum > FINDTALLY / 2}).

2. Ensure that the remediation deadline (\texttt{REMED\_DUE}) has already passed. 

3. Only change the audit status \texttt{AudtrailState} to \texttt{Overdue} if it is not already \texttt{Complete} or \texttt{Overdue}.

...

Here is the SQL query that meets the requirements:

\lstset{style=mystyle}
\begin{lstlisting}[language=SQL,
xleftmargin=20pt,
frame=tb,
framerule=0pt]
SELECT 
  AUDIT_TRACE, AudtrailState, FINDTALLY, ...
FROM 
  AuditAndCompliance
WHERE 
  AudtrailState NOT IN ('Complete', 'Overdue')
  AND critFindNum > FINDTALLY / 2
  AND REMED_DUE < CURRENT_DATE
\end{lstlisting}

However, the above query only filters the records. To actually change the status to \texttt{Overdue}, we need to perform an update operation.

\ldots

Finally, the correct SQL query is:

\vspace{1mm}
\noindent\textbf{Final output:}

\begin{lstlisting}[language=SQL,
xleftmargin=20pt,
frame=tb,
framerule=0pt]
SELECT * FROM AuditAndCompliance
WHERE 
  AudtrailState NOT IN ('Complete', 'Overdue')
  AND critFindNum > FINDTALLY / 2
  AND REMED_DUE < CURRENT_DATE
\end{lstlisting}
\end{mdframed}

\normalsize
\vspace{-2mm}
\caption{A case study illustrating RBAC failures in reasoning.
}
\label{fig:rbac_reasoning_failure}
\vspace{-3mm}
\end{figure}

As shown in Figure~\ref{fig:rbac_reasoning_failure}, the model explicitly reasons that the request requires modifying audit records by updating the audit status to \texttt{Overdue}.
This is a write operation on \texttt{AuditAndCompliance} table and should therefore be denied under restricted access.
However, despite identifying the need for an update operation during reasoning, the model ultimately produces a read-only \texttt{SELECT} query, effectively bypassing the access-control decision.

This reasoning-action inconsistency aligns with the \emph{refusal cliff} phenomenon~\cite{yin_refusal_2025}, in which a reasoning model maintains strong refusal intentions during internal reasoning but fails to preserve them in the final output.
Across RBAC instances that require denial, the model's chain-of-thought frequently reaches the correct \textsf{deny} conclusion, yet decoding favors a non-refusal SQL query.
This effect is more pronounced for harder queries with longer reasoning chains or more complex access constraints.
We hypothesize two potential contributing factors.
First, a \emph{decoding prior} toward task completion may outweigh refusal tokens during generation.
Second, \emph{post-training objectives} may emphasize usefulness or task success over strict policy adherence.
These factors suggest that enforcing RBAC compliance requires tighter coupling between reasoning, decoding, and policy-aware generation mechanisms.

\subsection{Limitations of Heuristic Remedies}
\label{subsec:heuristic_remedies}
Our analysis reveals significant safety gaps in current text-to-SQL systems, especially for open-weight models.
A natural question is whether these gaps can be mitigated using standard heuristic remedies.
To this end, we evaluate two commonly adopted approaches, \textit{supervised fine-tuning (SFT)} and \textit{in-context learning}.

\begin{table}[!t]
\centering
\footnotesize
\caption{RBAC supervised fine-tuning study.}
\vspace{-2mm}
\label{tab:rbac_spider_sft}
\setlength{\tabcolsep}{3.5pt}
\resizebox{\columnwidth}{!}{%
\begin{tabular}{@{}llrrrrr@{}}
\toprule
\textbf{Model} & \textbf{ } & \textbf{SafeEX(\%)$\uparrow$} & \textbf{SafeDeny(\%)$\uparrow$} & \textbf{Viol.(\%)$\downarrow$} & \textbf{OR(\%)$\downarrow$} & \textbf{AC-F1$\uparrow$} \\
\midrule

\multicolumn{1}{c}{} & \multicolumn{6}{c}{\textit{Evaluation on Spider-dev}} \\
\cmidrule(l){2-7}
\multirow{2}{*}{\makecell[l]{Llama3-SQL\\Coder-8b}}
& SFT     & 45.16$\pm$0.40 & 90.89$\pm$0.94 & 5.55$\pm$0.56 & 20.44$\pm$0.57 & 70.96$\pm$1.08 \\
& $\Delta$ & $-$15.37$\pm$1.03 & $+$90.89$\pm$0.94 & $-$44.85$\pm$0.95 & $+$20.44$\pm$0.57 & $+$4.66$\pm$0.97 \\
\cmidrule(l){2-7}

\multirow{2}{*}{\makecell[l]{Qwen2.5-\\Instruct-14b}}
& SFT     & 49.54$\pm$1.50 & 92.70$\pm$0.93 & 4.31$\pm$0.41 & 19.88$\pm$0.95 & 73.01$\pm$0.92 \\
& $\Delta$ & $-$19.20$\pm$1.64 & $+$66.35$\pm$3.15 & $-$32.63$\pm$2.06 & $+$19.34$\pm$1.06 & $+$0.17$\pm$1.45 \\

\midrule
\multicolumn{1}{c}{} & \multicolumn{6}{c}{\textit{Evaluation on BIRD-dev}} \\
\cmidrule(l){2-7}
\multirow{2}{*}{\makecell[l]{Llama3-SQL\\Coder-8b}}
& SFT     & 17.00$\pm$1.03 & 79.15$\pm$0.92 & 13.90$\pm$0.67 & 14.49$\pm$0.70 & 59.54$\pm$1.77 \\
& $\Delta$ & $-$21.21$\pm$1.72 & $+$79.07$\pm$0.89 & $-$52.29$\pm$0.95 & $+$14.49$\pm$0.70 & $+$9.04$\pm$1.56 \\
\cmidrule(l){2-7}

\multirow{2}{*}{\makecell[l]{Qwen2.5-\\Instruct-14b}}
& SFT     & 31.64$\pm$1.33 & 86.77$\pm$1.01 & 9.07$\pm$0.78 & 12.12$\pm$0.51 & 68.66$\pm$1.48 \\
& $\Delta$ & $-$13.50$\pm$1.50 & $+$71.56$\pm$0.59 & $-$47.12$\pm$0.69 & $+$11.99$\pm$0.52 & $+$14.02$\pm$1.15 \\

\midrule
\multicolumn{1}{c}{} & \multicolumn{6}{c}{\textit{Evaluation on LiveSQLBench}} \\
\cmidrule(l){2-7}
\multirow{2}{*}{\makecell[l]{Qwen2.5-\\Instruct-14b}}
& SFT     & 3.50$\pm$0.54 & 88.99$\pm$1.24 & 9.42$\pm$0.93 & 15.84$\pm$0.88 & 33.69$\pm$2.86 \\
& $\Delta$ & $-$7.02$\pm$2.03 & $+$61.21$\pm$1.33 & $-$50.98$\pm$1.37 & $+$14.73$\pm$0.96 & $-$2.12$\pm$2.56 \\
\bottomrule
\end{tabular}%
}
\end{table}

\balance
\vspace{1mm}
\noindent
\textbf{Supervised fine-tuning.} 
We construct a role-aware training dataset from the Spider training split using our role synthesis pipeline. 
Each instance is formatted as an instruction task containing the database schema, user query, and assigned role, with the target response being either the gold SQL for \textsf{allow} cases or a standardized refusal for \textsf{deny} cases. We evaluate two representative open-weight models, Llama-3-SQLCoder-8B and Qwen2.5-14B-Instruct, and defer implementation details to Appendix~\ref{appendix:sft-paras}.

\begin{figure}[!t]
\centering
\begin{minipage}[t]{0.49\linewidth}
    \centering
    \begin{tikzpicture}
        \begin{axis}[
            width=\linewidth,
            height=4.2cm,
            grid=major,
            grid style={dashed, gray!30},
            xlabel={\textbf{Safety (AC-F1)}},
            ylabel={\textbf{Utility (Safe-EX)}},
            xmin=30, xmax=100,
            ymin=0, ymax=80,
            tick label style={font=\footnotesize},
            label style={font=\footnotesize},
            title={(a) SFT Generalization},
            title style={
                font=\footnotesize,
                at={(0.5,-0.35)},
                anchor=north
            },
            legend style={font=\tiny, draw=none, fill=none},
        ]

        \draw [->, >=stealth, thick, blue!70!black] (axis cs: 66.30, 60.52) -- (axis cs: 70.96, 45.16);
        \addplot[mark=*, blue!70!black, mark size=1.5pt, forget plot] coordinates {(66.30, 60.52)};
        \addplot[mark=square*, blue!70!black, mark size=1.5pt, forget plot] coordinates {(70.96, 45.16)};

        \draw [->, >=stealth, thick, orange!80!black] (axis cs: 72.84, 68.74) -- (axis cs: 73.01, 49.54);
        \addplot[mark=*, orange!80!black, mark size=1.5pt, forget plot] coordinates {(72.84, 68.74)};
        \addplot[mark=square*, orange!80!black, mark size=1.5pt, forget plot] coordinates {(73.01, 49.54)};
        \node[anchor=west, font=\tiny, black, inner sep=1pt] at (axis cs: 78, 60) {Spider};

        \draw [->, >=stealth, thick, blue!70!black] (axis cs: 50.49, 38.21) -- (axis cs: 59.54, 17.00);
        \addplot[mark=*, blue!70!black, mark size=1.5pt, forget plot] coordinates {(50.49, 38.21)};
        \addplot[mark=square*, blue!70!black, mark size=1.5pt, forget plot] coordinates {(59.54, 17.00)};

        \draw [->, >=stealth, thick, orange!80!black] (axis cs: 54.64, 45.14) -- (axis cs: 68.66, 31.64);
        \addplot[mark=*, orange!80!black, mark size=1.5pt, forget plot] coordinates {(54.64, 45.14)};
        \addplot[mark=square*, orange!80!black, mark size=1.5pt, forget plot] coordinates {(68.66, 31.64)};
        \node[anchor=center, font=\tiny, black, inner sep=1pt] at (axis cs: 61, 33) {BIRD};

        \draw [->, >=stealth, thick, orange!80!black] (axis cs: 35.81, 10.51) -- (axis cs: 33.69, 3.50);
        \addplot[mark=*, orange!80!black, mark size=1.5pt, forget plot] coordinates {(35.81, 10.51)};
        \addplot[mark=square*, orange!80!black, mark size=1.5pt, forget plot] coordinates {(33.69, 3.50)};
        \node[anchor=west, font=\tiny, black, inner sep=1pt] at (axis cs: 44, 7) {LiveSQL};

        \node[anchor=north west, font=\tiny, align=left, inner sep=2pt,
              draw=gray!40, line width=0.3pt, rounded corners=1pt] at (axis cs: 31, 79) {%
            \textcolor{blue!70!black}{\rule[0.3ex]{6pt}{0.8pt}} Llama3 \\[-1pt]
            \textcolor{orange!80!black}{\rule[0.3ex]{6pt}{0.8pt}} Qwen
        };

        \node[anchor=south east, font=\tiny, align=left, inner sep=2pt,
              draw=gray!40, line width=0.3pt, rounded corners=1pt] at (axis cs: 99, 1) {%
            $\bullet$ ZS\quad \scalebox{0.65}{$\blacksquare$} SFT
        };

    \end{axis}
    \end{tikzpicture}
\end{minipage}
\hfill
\begin{minipage}[t]{0.49\linewidth}%
    \centering
    \begin{tikzpicture}
        \begin{axis}[
            width=\linewidth,
            height=4.2cm,
            xlabel={\textbf{Number of shots}},
            ylabel={\textbf{Safety (AC-F1)}},
            xtick={0, 2, 4, 6},
            ymin=0, ymax=100,
            grid=major,
            grid style={dashed, gray!30},
            tick label style={font=\footnotesize},
            label style={font=\footnotesize},
            title={(b) Prompting Efficacy},
            title style={
                font=\footnotesize,
                at={(0.5,-0.35)},
                anchor=north
            },
            legend style={
                at={(0.98, 0.05)},
                anchor=south east,
                font=\tiny,
                draw=none,
                fill=none,
                row sep=-2pt,
                nodes={inner sep=1pt},
                cells={anchor=west},
            },
        ]

        \addplot+[thick, color=blue!70!black, mark=*, mark options={scale=0.7}, error bars/.cd, y dir=both, y explicit]
            coordinates { (0, 31.90) +- (0, 0.92) (2, 36.15) +- (0, 2.88) (4, 34.88) +- (0, 2.66) (6, 35.75) +- (0, 2.72) };
        \addlegendentry{Gemma3}

        \addplot+[thick, color=orange!80!black, mark=square*, mark options={scale=0.7}, error bars/.cd, y dir=both, y explicit]
            coordinates { (0, 37.28) +- (0, 1.82) (2, 31.33) +- (0, 2.23) (4, 30.47) +- (0, 0.93) (6, 30.44) +- (0, 1.89) };
        \addlegendentry{DeepSeek}

        \addplot+[thick, color=red!70!black, mark=triangle*, mark options={scale=0.9}, error bars/.cd, y dir=both, y explicit]
            coordinates { (0, 58.25) +- (0, 2.77) (2, 58.62) +- (0, 2.77) (4, 56.44) +- (0, 3.30) (6, 58.11) +- (0, 2.96) };
        \addlegendentry{GPT-5-mini}

        \end{axis}
    \end{tikzpicture}
    \label{subfig:few-shot}
\end{minipage}

\vspace{-3mm}
\caption{\textbf{Analysis of heuristic remedies.}}
\label{fig:mitigation_analysis}
\end{figure}
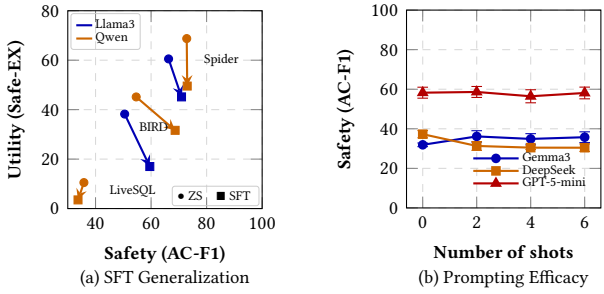

Table~\ref{tab:rbac_spider_sft} and Figure~\ref{fig:mitigation_analysis}(a) compare fine-tuned models with their zero-shot baselines.
On the in-domain RBAC-Spider benchmark, SFT substantially improves safety compliance.
For example, Llama3-SQLCoder-8B improves its AC-F1 from 66.3 to 71.0, with Safe-Deny rising from near zero to over $90\%$ and violation rate dropping from $50\%$ to $6\%$.
This shows that SFT can enforce RBAC in familiar domains. 
However, this safety gain comes at a clear utility cost, as Safe-EX drops on both models due to elevated over-refusal.

We next evaluate whether these improvements generalize to the more complex RBAC-BIRD and RBAC-LiveSQLBench benchmarks.
The limitations of fine-tuning become apparent.
Although AC-F1 improves on BIRD ($+9.0$ for Llama3-SQLCoder and $+14.0$ for Qwen2.5-14B), it drops on LiveSQLBench ($-2.1$ for Qwen2.5-14B), and the fine-tuned models remain well below top commercial models such as GPT-5 across all out-of-domain settings.
Moreover, Safe-EX drops sharply and Over-Refusal rate increases, while Safe-Deny remains high ($>$70\%).
This indicates that the fine-tuned models become biased toward refusal rather than learning generalized RBAC reasoning.
Instead of acquiring a generalized understanding of access control, the LLM adopts an ineffective, risk-averse denial strategy.
Thus, text-to-SQL under RBAC constraints remains a complex task, not merely a pattern-matching problem solvable by fine-tuning on a single dataset.

\vspace{1mm}
\noindent\textbf{In-context learning.} 
We further examine whether few-shot prompting can improve RBAC compliance without updating model parameters.
We conduct an ablation study on the challenging RBAC-LiveSQLBench using prompts with $k \in \{2,4,6\}$ demonstrations, in addition to the zero-shot baseline.
To avoid biasing decisions toward refusal or acceptance, we curate demonstrations in \textit{balanced pairs}, where each additional two shots adds one authorized and one unauthorized query.
All experiments follow the stratified multi-pass evaluation protocol in Section~\ref{sec:metrics_and_tools:evaluation_tools} for fair comparison.

Figure~\ref{fig:mitigation_analysis}(b) shows that few-shot prompting does not provide consistent safety improvements.
Contrary to the expectation that more context improves compliance, we observe \textit{non-monotonic fluctuations} across models.
While balanced demonstrations help Gemma3-27B, improving AC-F1 from {32\%} to {36\%}, they cause a safety regression in DeepSeek-Coder, where AC-F1 drops by nearly {6} points after receiving examples.
This suggests that, for some models, few-shot examples introduce noise or distract from intrinsic refusal boundaries rather than clarifying RBAC logic.
For stronger models such as GPT-5-mini, additional demonstrations yield negligible gains, indicating a performance ceiling.
Consequently, few-shot prompting is highly model-dependent and unstable, making it insufficient for security-critical database interfaces.

\subsection{Implications for Practical Deployment}\label{sec:exp:implications}
Our experiments reveal a fundamental challenge for practitioners deploying text-to-SQL systems in access-controlled DBMS environments: such systems must simultaneously achieve high SQL generation accuracy and reliable RBAC compliance.
The SFT results indicate that fine-tuning open-weight models can substantially improve compliance in a controlled setting.
However, this approach is relatively brittle and does not readily extend to more complex application scenarios.
As shown in the previous subsection, safety alignment learned on Spider does not generalize to BIRD or LiveSQLBench, implying that fine-tuning may be required separately for each database.
In practice, this may further require maintaining distinct LoRA adapters for different user roles, similar to security-domain isolated deployments~\cite{jayaraman_permissioned_2025}, which is difficult to scale and costly to maintain in real systems.

Leading commercial models, in contrast, show stronger zero-shot adherence to RBAC constraints, reflecting more generalizable instruction-following capabilities.
These models provide a higher baseline level of access-control compliance, likely due to extensive safety alignment during post-training.
Table~\ref{tab:overall} reports the mean per-query cost for role-free and role-aware executions, illustrating the operational cost of such models.
Per-task costs are estimated from input/output token counts using each provider's published pricing; open-weight models are priced via DeepInfra's hosted inference rates where available, and reported as ``--'' otherwise.
This gives database practitioners a practical trade-off between the control and transparency of open-weight fine-tuning and the convenience and stronger default safety of higher-cost commercial models.

Overall, reliable RBAC is not solved by model selection alone.
Improving the intrinsic safety of the LLM is necessary, but robust deployment must also combine language reasoning with external, deterministic access-control enforcement.
Our benchmark provides a practical framework for measuring this progress and comparing trade-offs among safety, utility, and cost in enterprise-oriented natural language database interfaces.

\section{Related Work}\label{sec:related_work}

\noindent \textbf{Text-to-SQL benchmarks and methods.} Text-to-SQL has evolved from a semantic parsing task into a core database application~\cite{li2014nalir,li2023tablegpttabletunedgptdiverse, urban2023caesuralanguagemodelsmultimodal,zhang_finsql_2024,luoma2025snails,yang2025automated,xie2025opensearch,chen2025reliable,li2024thedawn,fu2023catsql,gu2023few}, driven by increasingly challenging benchmarks. 
Spider~\cite{yu_spider_2019} established cross-domain generalization over complex schemas, while later benchmarks such as BIRD~\cite{li_can_2023} introduced enterprise-scale databases with noisy and domain-specific data. 
More recently, LiveSQLBench~\cite{livesqlbench_hf} simulates interactive analytical workflows, and EvoSchema~\cite{zhang2025evoschema} studies robustness to schema evolution. Despite advancing text-to-SQL accuracy and efficiency~\cite{li2024thedawn}, these benchmarks largely assume unrestricted database access and do not consider access control constraints that are ubiquitous in practical DBMS environments. 

Accordingly, recent text-to-SQL methods focus on supporting complex SQL generation through modular pipelines~\cite{hong_next-generation_2025, pourreza2023dinsqldecomposedincontextlearning,fan2024combining}. 
These include in-context reasoning with task decomposition (e.g., DIN-SQL~\cite{pourreza2023dinsqldecomposedincontextlearning}), database-oriented pre-processing such as schema pruning and example selection~\cite{ren2024purple, gao_text--sql_2024}, domain-specific fine-tuning via instruction tuning or multi-task learning~\cite{li2024codes, chang-fosler-lussier-2023-selective, xie2022unifiedskgunifyingmultitaskingstructured}, and post-generation refinement using execution-based verification~\cite{fan2024metasql}. 
Despite these advances, existing work has almost exclusively targeted semantic correctness under full-access assumptions, leaving access control compliance largely unexamined.

\vspace{1mm}
\noindent \textbf{Database access control.} 
Secure access in enterprise databases is traditionally governed by established access control models, including discretionary and mandatory access control~\cite{485845, ferraiolo2003role}, with role-based access control (RBAC) being the dominant paradigm in practice.
RBAC assigns permissions to roles rather than users, enabling scalable governance, but integrating LLM-based query generation introduces new challenges, as access decisions must interact with probabilistic model outputs.

Recent work has begun to explore how LLM-based systems can be integrated with access control mechanisms. 
Database-side approaches, such as BridgeScope~\cite{weng_bridgescope_2025}, enforce privileges via tool-based execution but often expose full schemas to preserve SQL accuracy, increasing the risk of schema inference or prompt-based attacks~\cite{shay2019querycontrol,luo2025prompt}. 
Parameter-based approaches, including PermLLM~\cite{jayaraman_permissioned_2025} and related adapter-based methods~\cite{huang_lorahub_2024, saha2025sudollmmultirolealignmentlanguage, fleshman2025adapterswapcontinuoustrainingllms, almheiri_role-aware_2025}, embed access restrictions into model parameters, but typically assume clearer domain boundaries and do not generalize well to dynamic, cross-domain SQL generation. 
A concurrent work~\cite{klisura2025role} examines role-conditioned refusal on Spider and BIRD using manually specified roles and read-only queries. 
In contrast, our work introduces an automated framework for synthesizing realistic roles and fine-grained RBAC policies for arbitrary text-to-SQL benchmarks, and evaluates compliance on LiveSQLBench with full CRUD workloads. We further conduct a large-scale empirical study on this more realistic RBAC setting, enabling systematic analysis of failure modes that are not visible under existing benchmarks.

These trade-offs highlight a severe tension between security, performance, and flexibility. Simple mitigations, such as restricting schema exposure based on roles, are insufficient, as we demonstrate empirically in Section~\ref{subsec:schema_expose}. Other efforts, including OrgAccess~\cite{sanyal_orgaccess_2025} and DePLOI~\cite{subramaniam_deploi_2025}, address complementary problems, such as natural language reasoning over access control policies or using text-to-SQL to synthesize and audit policies. 
However, they do not address the core challenge of 
evaluating whether a text-to-SQL model can generate or abstain according to role policies at inference time, while final access enforcement remains deterministic, 
which highlights the need for a dedicated benchmark.

\balance

\section{Conclusion}\label{sec:conclusion}

In this paper, we introduced a benchmarking framework for evaluating text-to-SQL systems under RBAC constraints. The framework augments existing text-to-SQL benchmarks with realistic roles, fine-grained access policies, RBAC-aware ground truth, and compliance-aware metrics. We instantiate it on three widely used datasets, producing RBAC-augmented benchmarks for systematic evaluation.
Our empirical study shows that current text-to-SQL systems, despite strong execution accuracy under unrestricted access, often violate RBAC policies when access constraints are enforced.
Common mitigation strategies, including schema restriction, prompt-based policy specification, and supervised fine-tuning, do not reliably eliminate these failures.

These results show a clear gap between existing text-to-SQL evaluation and the needs of access-controlled database deployments.
They also motivate further study of text-to-SQL under RBAC constraints, which remains an important but underexplored problem for practical database systems.
Future work includes richer role hierarchies and more varied access policy patterns, such as instance-level access control, which are not covered by current benchmarks.
Our benchmark focuses on \emph{explicit} RBAC compliance, i.e., whether generated SQL references only authorized resources.
It does not model inference-based leakage, such as deriving restricted attributes from correlated authorized ones.
Capturing such indirect channels requires policy models beyond standard RBAC, such as inference control or semantic privacy, and is left for future work.

\balance
\clearpage

\bibliographystyle{ACM-Reference-Format}
\bibliography{refs}

\clearpage

\appendix
\nobalance
\section{Details of Human-in-the-Loop Validation}\label{appendix:human-eval}
\begin{figure}[htbp]
    \centering
        \includegraphics[width=1.2\linewidth, trim=0.8cm 0.8cm 0cm 0.5cm, clip]{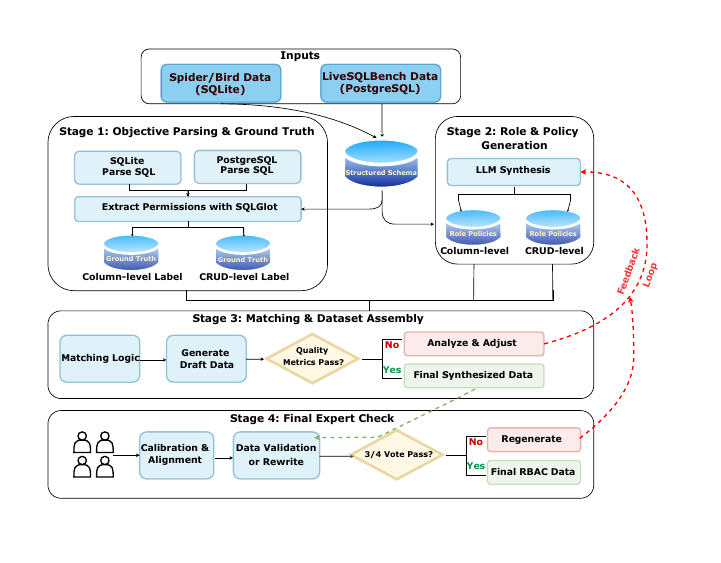}
    \vspace{-5mm}
    \caption{Pipeline of data generation with human in the loop.}
    \label{fig:human-validation}
\end{figure}

This section provides the full details of the human-in-the-loop validation procedure briefly described in Section~\ref{sec:data_gen:generation}. 
An overview of the complete validation pipeline is illustrated in Figure~\ref{fig:human-validation}.

\subsection{Automatic Quality Metrics}
Before any human evaluation, each synthesized role configuration undergoes a deterministic and fully automated quality screening process. 
These checks apply to the LLM-synthesized semantic roles; the scoped DataOperator roles are generated programmatically rather than by LLM synthesis and therefore do not undergo this semantic-role screening.

Given a database schema and its synthesized role set, we compute the following interpretable metrics.

\begin{enumerate}[leftmargin=*]
    \item \underline{\textit{Denial rate}}. The denial rate measures the fraction of role-query pairs that are labeled as \textsf{deny} under the synthesized policies:
    \begin{align*}
        \text{DenyRate}=\frac{\#\{\textsf{deny}\}}{\#\{\textsf{total role-query pairs}\}}.
    \end{align*}
    This metric is used to detect configurations that are overly permissive (very low denial rate $\leq 5\%$) or overly restrictive (very high denial rate $\geq90\%$), both of which reduce the usefulness of the benchmark. 
    \item \textit{\underline{Coverage standard deviation}}. To measure how access permissions are distributed across roles, we compute coverage ratios for each role $r$ and each SQL operation type. For \texttt{INSERT} and \texttt{DELETE}, coverage is measured at the table level:
    \[\text{Coverage}_\texttt{INSERT}(r)=\frac{|\text{tables authorized for \texttt{INSERT} by $r$}|}{|\text{total tables}|},\]
    and analogously for \texttt{DELETE}.
    For \texttt{SELECT} and \texttt{UPDATE}, coverage is measured at the column level:
    \[\text{Coverage}_\texttt{SELECT}(r)=\frac{|\text{columns authorized for \texttt{SELECT} by $r$}|}{|\text{total columns}|},\]
    and analogously for \texttt{UPDATE}. We then compute the standard deviation of coverage values across roles for each operation type.
    For LiveSQLBench, these standard deviations are further combined using operation-frequency weights derived from the empirical query distribution, whereas for Spider and BIRD we directly use the \texttt{SELECT} coverage standard deviation.
    Configurations with extremely low ($\leq0.1$) coverage variance indicate nearly identical roles, and are therefore rejected. 
    \item \textit{\underline{Policy overlap}}. To quantify redundancy among roles, we compute pairwise Jaccard similarity between role permission sets:
    \[\text{Overlap}(r_i,r_j)=\frac{|P_i\cap P_j|}{|P_i\cup P_j|},\]
    where $P_i$ denotes the set of permissions associated with role $r_i$.
    For each role configuration, we consider the maximum overlap across all role pairs.
    If this maximum exceeds a fixed threshold (0.8 in our implementation), the configuration is rejected, as it indicates insufficient role distinctness.
    \item \textit{\underline{Semantic similarity}}. To assess whether synthesized roles are semantically grounded in the database schema, we compute cosine similarity between embeddings of role descriptions and the schema description.
    Specifically, role names and descriptions are concatenated and embedded, and compared against the embedded schema text.
    We use 0.60 as an auxiliary threshold; human experts make the final semantic assessment.
\end{enumerate}

All metric computations are deterministic and produce explicit numeric values and pass/fail signals. 
Once a role configuration fails any automatic check, the violated metrics and their values are packaged as structured feedback and supplied to the role synthesis stage for regeneration, as illustrated in Figure~\ref{fig:human-validation}.

\subsection{Reject-Regenerate Feedback Loop}
When a synthesized role configuration fails one or more automatic quality checks, it is rejected and returned to the role synthesis stage.
The rejection feedback is packaged as structured input to the LLM and includes:
(i) which quality metrics were violated;
(ii) the corresponding metric values; and
(iii) a short explanation of what each violated metric indicates (e.g., ``roles are overly similar,'' ``access is too restrictive'').
The context-grounded role and policy generation process is then re-executed with this augmented feedback.
This reject-regenerate loop iterates until satisfying all automatic quality.

\subsection{Human Validation Protocol}
Once a role configuration passes all automatic checks, it is subjected to final human validation.
Human annotators do not label individual role-query pairs; instead, they review each role configuration at the database level.

\subsubsection{Annotator Setup} Human validation is conducted by four annotators with prior expertise in databases and access control. 
Annotators are provided with the database schema, synthesized role names and descriptions, and a summarized view of each role's access scope.
Before annotation, annotators undergo a light-weight calibration on a held-out subset to align on evaluation criteria.

\subsubsection{Validation Criteria} 
Annotators evaluate each RBAC role configuration along the following dimensions: (i) whether role definitions correspond to plausible organizational responsibilities implied by the schema; (ii) whether access scopes are coherent with role semantics; (iii) whether the overall role set reflects reasonable overlap and separation of responsibilities. 
Annotators provide a binary accept/reject decision for each configuration.
In cases of rejection, annotators provide a high-level reason (e.g. ``roles lack clear functional distinction,'' ``permissions do not match role semantics'').

\subsubsection{Disagreement Resolution}
A role configuration is accepted if and only if it receives a majority approval (at least three out of four accept votes).
If the role configuration fails to reach majority approval, it is rejected.
For rejected cases, the collected rejection reasons are consolidated and returned to the synthesis pipeline as structured feedback, and the configuration is regenerated following the same reject-regenerate procedure described above.
Only role configurations that pass both automatic screening and human validation are retained for subsequent permission verification and dataset construction.

\subsubsection{Statistics.} \label{appendix:human-validate-stats}
In the final construction, 28 of 53 database-level role configurations passed validation on the first attempt; 16 were accepted after one feedback-guided regeneration round; and 9 were directly revised by annotators. 
No configuration required more than one regeneration round and the process took 4 working days.

\section{Details in Generating DataOperators}
\label{appendix:gen-do}

The \texttt{DataOperator} roles complement the semantically synthesized roles by serving as scoped administrators, while each individual role still exposes meaningful access restrictions.

Algorithm~\ref{alg:do-generation} formalizes the procedure: for each database, we instantiate $n = 3$ roles ($n = 2$ when the database has fewer than three tables), and sample each role $\pi_i$ \emph{independently per table} (random seed = 42), with probability $p_{\mathit{star}} = 0.5$ the role receives full-table access (\texttt{*}); otherwise, each column is independently included with probability $p_{\mathit{col}} = 0.7$, subject to a minimum of one column and a cap of $\lfloor 0.9 \cdot |\mathit{cols}(T)| \rfloor$ columns. These parameters balance the two access granularities and give each column a marginal coverage of $0.85$ per role. Since the procedure is probabilistic, we empirically verify column coverage on the final benchmark: the role union achieves $100\%$ coverage on every Spider (20/20) and BIRD (11/11) database, and on 18/22 LiveSQLBench databases ($99.1\%$ on average), with any residual uncovered columns treated as universally denied by construction.
For LiveSQLBench, \texttt{SELECT} and \texttt{UPDATE} share the same column scope, \texttt{INSERT} and \texttt{DELETE} are table-level over the accessible tables.

\begin{algorithm}[htbp]
  \caption{DataOperator Policy Generation}
  \label{alg:do-generation}
  \small
  \KwIn{Schema $\mathcal{S} = \{T_1, T_2, \dots\}$ where each table $T$ has columns $\mathit{cols}(T)$;
  parameters $p_{\mathit{star}}{=}0.5$, $p_{\mathit{col}}{=}0.7$, $\mathit{cap}{=}0.9$.}
  \KwOut{Set of \texttt{DataOperator} policies $\{\pi_1, \dots, \pi_n\}$.}

  $n \gets 3$ \textbf{if} $|\mathcal{S}| \geq 3$ \textbf{else} $2$\;
  \For{$i \gets 1$ \KwTo $n$}{
    $\pi_i \gets \emptyset$\;
    \ForEach{table $T \in \mathcal{S}$}{
      \uIf{$\mathrm{rand}() < p_{\mathit{star}}$}{
        $\pi_i[T] \gets \{\texttt{*}\}$ \tcp*{full-table access}
      }
      \Else{
        $C \gets \{\, c \in \mathit{cols}(T) \mid \mathrm{rand}() < p_{\mathit{col}} \,\}$\;
        \lIf{$C = \emptyset$}{$C \gets \{\mathrm{random\_choice}(\mathit{cols}(T))\}$}
        \lIf{$|C| > \lfloor \mathit{cap} \cdot |\mathit{cols}(T)| \rfloor$}{$C \gets \mathrm{sample}(C, \lfloor \mathit{cap} \cdot |\mathit{cols}(T)|\rfloor)$}
        $\pi_i[T] \gets C$\;
      }
    }
  }
  \Return $\{\pi_1, \dots, \pi_n\}$\;
\end{algorithm}
\section{Parameters and Detailed SFT Settings}
\label{appendix:sft-paras}

The fine-tuning was implemented using a modified version of the {DB-GPT-Hub} (\url{https://github.com/eosphoros-ai/DB-GPT-Hub/})  framework. We fine-tuned two widely adopted instruction models as listed in Table~\ref{tab:sft_models}.

We performed supervised fine-tuning (SFT) using a standard causal language modeling objective on the instruction-formatted dataset. During training, prompt tokens were masked from the cross-entropy loss calculation to ensure that optimization occurred exclusively on the target assistant responses. For computational efficiency, we employed the QLoRA technique for parameter-efficient fine-tuning (PEFT). This involved loading the base models in 4-bit precision using \texttt{BitsAndBytesConfig} and injecting LoRA adapters. A PEFT-aware trainer was then used to optimize only the low-rank adapter parameters, keeping the original model weights frozen. Detailed parameters are shown in Table~\ref{tab:sft_hparams}.
To ensure DeepSpeed multi-process training exhibits identical random behavior, 
we set a global random seed equals to 42 at the beginning of the Trainer initialization.

\begin{table}[!t]
\centering
\small
\caption{Base models for fine-tuning experiments. LoRA targets are set default as q\_proj and v\_proj.}
\label{tab:sft_models}
\begin{tabular}{@{}ll@{}}
\toprule
\textbf{Model Name} & \textbf{HuggingFace Identifier} \\
\midrule
Llama-3-SQLCoder-8b & \texttt{defog/llama-3-sqlcoder-8b} \\
Qwen2.5-14B-Instruct & \texttt{Qwen/Qwen2.5-14B-Instruct} \\
\bottomrule
\end{tabular}
\end{table}

\begin{table}[!t]
\centering
\small
\caption{Hyperparameters for QLoRA fine-tuning.}
\vspace{-2mm}
\label{tab:sft_hparams}
\setlength{\tabcolsep}{4pt}
\begin{tabular}{@{}llll@{}}
\toprule
\textbf{Parameter} & \textbf{Value} & \textbf{Parameter} & \textbf{Value} \\
\midrule
Finetuning Type & QLoRA   & LoRA Alpha        & 32   \\
Quantization    & 4-bit   & Learning Rate     & 2e-4 \\
Precision       & bf16    & Epochs            & 8    \\
LoRA Rank       & 64      & Max Source Length & 2048 \\
\bottomrule
\end{tabular}
\end{table}

\section{Row-Level RBAC Feasibility Study}
\label{appendix:row-level}

We conduct a limited feasibility study to examine whether the framework introduced in Section~\ref{sec:data_generation} can be extended to row-level RBAC.
We select five Spider databases containing row-discriminative attributes (e.g., department, region, year) and augment each role policy with a row-level predicate.
The role synthesis process is reused, while policy verification is extended to check whether the predicate induced by the gold SQL is compatible with the role's permitted row scope.
This yields 298 (query, role) instances expanded from 134 questions.
We evaluate four representative models (two commercial, two open-weight) using the same fair-comparison (5 trials, random role sampling), results are reported in Table~\ref{tab:spider_row_level_model_comparison}.

\begin{table}[htbp]
  \centering
  \footnotesize
  \setlength{\tabcolsep}{5pt}
  \caption{Safety performance on Spider Row-level.}
  \vspace{-2mm}
  \label{tab:spider_row_level_model_comparison}
  \begin{tabular}{lrcc}
    \toprule
    \textbf{Model} & \textbf{Viol-Rate(\%)}$\downarrow$ & \textbf{OR-Rate(\%)}$\downarrow$ & \textbf{AC-F1}$\uparrow$ \\
    \midrule
    DeepSeek-Coder & 50.34 $\pm$ 1.57 & 0.94 $\pm$ 0.25 & 56.63 $\pm$ 2.24 \\
    Claude-Sonnet-4.5 & 2.82 $\pm$ 0.76 & 9.93 $\pm$ 1.61 & 79.35 $\pm$ 2.67 \\
    GPT-4o-mini & 25.97 $\pm$ 1.28 & 5.64 $\pm$ 1.26 & 64.57 $\pm$ 2.22 \\
    GPT-5 & 2.15 $\pm$ 0.66 & 9.19 $\pm$ 1.71 & 81.71 $\pm$ 3.46 \\
    \bottomrule
  \end{tabular}
  \vspace{-2mm}
\end{table}

The results show that RBAC compliance errors remain nontrivial under row-level predicates.
Commercial models achieve higher AC-F1 and lower violation rates than open-weight models.
This suggests that our construction and evaluation methodology extends to finer-grained policies, although we do not treat row-level RBAC as a full benchmark track in this work.
Scaling it to comparable size would require additional database selection, predicate engineering, and human auditing, which we leave to future work.

\vspace{-2mm}

\section{Per-Difficulty Results}
\label{appendix:per-difficulty}

We investigate the relationship between text-to-SQL task complexity and the effectiveness of RBAC enforcement.

\begin{figure}[htbp]
  \centering
  \small
  \setlength{\fboxsep}{0pt}

  \begin{tcolorbox}[colback=black!2,colframe=black!35,boxrule=0.45pt,
                    sharp corners, left=4pt,right=4pt,top=4pt,bottom=4pt,
                    width=1.0\columnwidth]
    \textbf{Input}: "I want to identify which primary diagnoses are associated with the highest Crisis Intervention Frequency (CIF) across all patients... (omitted for brevity) ...Sort the results by CIF in descending order.",
  \end{tcolorbox}
  
  \begin{tcolorbox}[colback=black!2,colframe=black!35,boxrule=0.45pt,
                    sharp corners, left=4pt,right=4pt,top=4pt,bottom=4pt,
                    width=1.0\columnwidth]
    \textbf{Step A: Latest diagnosis per patient}\\
    \textbf{Idea:} pick latest \texttt{primdx} per patient via window function.\\
    \textbf{Required Schema:} \\
    \textsf{assessmentsocialanddiagnosis}(\texttt{primdx}, \texttt{patownerref}),\\
    \textsf{encounters}(\texttt{timemark}).
    \tcbline
    \textbf{Step B: Crisis interventions per patient}\\
    \textbf{Idea:} sum \texttt{crisisint} by \texttt{patref}.\\
    \textbf{Required Schema:} \\
    \textsf{treatmentbasics}(\textcolor{red!70}{\textbf{\texttt{crisisint}}}, \texttt{patref}).
    \tcbline
    \textbf{Step C: Total patients \& join}\\
    \textbf{Idea:} join A+B and compute CIF.\\
    \textbf{Constraint Check:} Verify column access for all steps.
  \end{tcolorbox}

  \begin{tcolorbox}[colback=black!2,colframe=black!35,boxrule=0.45pt,
                    sharp corners, left=4pt,right=4pt,top=4pt,bottom=4pt,
                    width=1.0\columnwidth]
    \textbf{Column-Level Role Policy \(\Pi_r\)}\\
    \emph{\instancecolorbox{green!15}{Allowed Columns:}} \\
    \textsf{assessmentsocialanddiagnosis}.*, \\
    \textsf{encounters}.\{\texttt{timemark}, \texttt{patownerref}\}, \\
    \textsf{treatmentbasics}.\{\texttt{patref}, \texttt{treatdate}\}. \\
    \emph{\instancecolorbox{red!15}{Denied Columns:}}     \textsf{treatmentbasics}\textcolor{red!70}{\textbf{\texttt{(crisisint)}}} \textit{(Sensitive clinical data)}.
    \tcbline
    \textbf{Access Control:} \\ 
    if any required column \(\not\subseteq \Pi_r\) $\rightarrow$ ``Sorry, I cannot answer''.
  \end{tcolorbox}
  \vspace{-4mm}
  \caption{A concrete challenge-level RBAC \texttt{SELECT} task.}
  \label{fig:gold-sql-rbac}
  \vspace{-3mm}
\end{figure}

\begingroup
\begin{table}[htbp]
\caption{RBAC-LiveSQLBench per-category performance.}
\vspace{-2mm}
\label{tab:livesql-singlecol}
\centering
\scriptsize
\setlength{\tabcolsep}{4pt}
\begin{adjustbox}{width=1.06\columnwidth,center}
\begin{tabulary}{\columnwidth}{L C C R R R C}
\toprule
\textbf{Model} & \textbf{Category} & \textbf{EX} & \textbf{Safe-EX} & \textbf{Viol. (\%)} & \textbf{OR (\%)} & \textbf{AC-F1} \\
\midrule

\multirow{2}{*}{\makecell[l]{Snowflake- \\ R1-7b}}
  & Query        & 5.61 & 5.97$\pm$2.50 & 84.54$\pm$1.04 & 0.00$\pm$0.00 & 26.71$\pm$1.51 \\
  & Management  & 6.59 & 16.57$\pm$4.04 & 76.37$\pm$2.30 & 0.88$\pm$0.27 & 35.03$\pm$3.01 \\
\cmidrule(lr){2-7}

\multirow{2}{*}{\makecell[l]{Gemma3- \\ 4b}}
  & Query        & 2.44 & 2.98$\pm$1.06 & 79.80$\pm$0.85 & 1.76$\pm$0.42 & 24.54$\pm$1.16 \\
  & Management  & 5.49 & 9.23$\pm$2.97 & 57.25$\pm$1.95 & 4.40$\pm$1.25 & 39.31$\pm$3.37 \\
\cmidrule(lr){2-7}

\multirow{2}{*}{\makecell[l]{Gemma3- \\ 27b}}
  & Query        & 6.59 & 7.55$\pm$1.51 & 66.88$\pm$0.78 & 1.32$\pm$0.20 & 28.81$\pm$1.01 \\
  & Management  & 12.64 & 13.56$\pm$3.45 & 45.93$\pm$2.27 & 3.30$\pm$0.98 & 46.56$\pm$1.89 \\
\cmidrule(lr){2-7}

\multirow{2}{*}{\makecell[l]{Qwen2.5- \\ Coder-7b}}
  & Query        & 3.41 & 5.11$\pm$0.96 & 80.63$\pm$1.75 & 0.39$\pm$0.29 & 26.83$\pm$2.59 \\
  & Management  & 8.79 & 13.72$\pm$2.04 & 56.37$\pm$1.28 & 1.87$\pm$0.82 & 43.91$\pm$2.30 \\
\cmidrule(lr){2-7}

\multirow{2}{*}{\makecell[l]{Qwen2.5- \\ 14b}}
  & Query        & 6.83 & 8.06$\pm$2.42 & 69.22$\pm$1.41 & 1.07$\pm$0.33 & 28.76$\pm$1.59 \\
  & Management  & 8.24 & 13.41$\pm$1.99 & 37.36$\pm$2.93 & 1.21$\pm$0.41 & 58.33$\pm$3.13 \\
\cmidrule(lr){2-7}

\multirow{2}{*}{\makecell[l]{Deepseek- \\ Reasoner}}
  & Query        & 25.00 & 25.64$\pm$3.64 & 30.39$\pm$1.22 & 3.27$\pm$0.57 & 46.16$\pm$1.08 \\
  & Management  & 18.75 & 17.85$\pm$6.32 & 18.79$\pm$2.98 & 12.53$\pm$1.41 & 54.20$\pm$4.44 \\
\cmidrule(lr){2-7}

\multirow{2}{*}{\makecell[l]{Deepseek- \\ Coder}}
  & Query        & 20.19 & 20.45$\pm$2.15 & 51.61$\pm$2.43 & 2.44$\pm$0.64 & 31.95$\pm$3.00 \\
  & Management  & 15.91 & 18.47$\pm$2.45 & 30.66$\pm$1.89 & 3.63$\pm$1.08 & 54.09$\pm$1.56 \\
\cmidrule(lr){2-7}

\multirow{2}{*}{\makecell[l]{Claude- \\ Sonnet-4.5}}
  & Query        & 24.63 & 22.54$\pm$3.88 & 16.24$\pm$2.32 & 4.68$\pm$0.73 & 55.81$\pm$4.18 \\
  & Management  & 15.38 & 13.47$\pm$6.45 & 6.59$\pm$1.25 & 17.36$\pm$1.92 & 55.73$\pm$3.27 \\
\cmidrule(lr){2-7}

\multirow{2}{*}{\makecell[l]{Gemini- \\ 2.5-Flash}}
  & Query        & 23.90 & 16.92$\pm$3.29 & 43.46$\pm$2.56 & 1.41$\pm$0.39 & 39.49$\pm$2.47 \\
  & Management  & 15.38 & 15.47$\pm$7.33 & 16.92$\pm$1.79 & 9.12$\pm$3.18 & 58.99$\pm$3.62 \\
\cmidrule(lr){2-7}

\multirow{2}{*}{\makecell[l]{GPT- \\ 4o-mini}}
  & Query        & 12.50 & 7.33$\pm$2.71 & 30.97$\pm$0.80 & 8.24$\pm$0.94 & 29.16$\pm$2.86 \\
  & Management  & 14.20 & 8.99$\pm$2.17 & 12.75$\pm$1.78 & 21.21$\pm$3.29 & 33.51$\pm$3.78 \\
\cmidrule(lr){2-7}

\multirow{2}{*}{\makecell[l]{GPT- \\ 5-mini}}
  & Query        & 20.24 & 26.36$\pm$1.85 & 23.13$\pm$2.11 & 2.49$\pm$0.32 & 53.86$\pm$2.46 \\
  & Management  & 14.84 & 24.67$\pm$1.89 & 11.98$\pm$0.81 & 8.46$\pm$1.08 & 67.63$\pm$2.18 \\
\cmidrule(lr){2-7}

\multirow{2}{*}{\makecell[l]{GPT-5}}
  & Query        & 25.12 & 28.70$\pm$3.75 & 20.54$\pm$2.38 & 2.54$\pm$0.53 & 57.27$\pm$3.66 \\
  & Management  & 31.87 & 28.64$\pm$2.43 & 11.76$\pm$1.73 & 9.01$\pm$2.73 & 68.41$\pm$2.25 \\

\bottomrule
\end{tabulary}
\end{adjustbox}
\end{table}
\endgroup

Difficulty annotations are present for both Spider and BIRD. Spider categorized task difficulty into four levels: easy, medium, hard, and extra hard, based on the complexity of SQL components, such as the number of \texttt{SELECT} columns, \texttt{WHERE} conditions, use of \texttt{GROUP BY}, nested subqueries, and advanced operations like \texttt{EXCEPT} or \texttt{INTERSECT}. BIRD provides a difficulty label that includes simple, moderate, and challenge in their dataset. 
However, LiveSQLBench only provides operation labels as Query or Management rather than difficulty labels, we follow this taxonomy as another distinguishing way between them. Figure~\ref{fig:gold-sql-rbac} shows a concrete challenge-level task, on which almost all models commit RBAC violations.

Table~\ref{tab:livesql-singlecol} shows that in RBAC-LiveSQLBench, as task category changes, text-to-SQL models struggle more with access control compliance. Specifically, we observe that AC-F1 score consistently decreases from management to more complex query tasks across all models. This decline indicates that models become more prone to access control errors as task difficulty increases. A closer look reveals for most LLMs, RBAC violations are more common than over-refusals, which suggests that models prioritize generating executable SQL queries over adhering to access control constraints.

The degradation in access control performance is most pronounced on LiveSQLBench, where nearly all models, particularly open-weight ones, exhibit exceptionally high violation rates. For instance, 
{Gemma-3-27B's violation rate surges to 60.44\%}
, a sharp increase from its performance on Spider and BIRD. This is likely attributable to the benchmark's design, which mirrors complex enterprise environments with long and detailed schema descriptions. The extensive context can overwhelm the model's attention mechanisms, distracting it from the specific RBAC constraints outlined in the prompt. Table~\ref{tab:bird-singlecol} shows the per-difficulty results for all tested models on BIRD while Table~\ref{tab:spider-singlecol} shows per-difficulty results on Spider, which lead to similar conclusions.

\begin{table*}[htbp]
\caption{RBAC-BIRD difficulty tier performance.}
\vspace{-2mm}
\centering
\small
\setlength{\tabcolsep}{4pt}
\begin{adjustbox}{max totalsize={\textwidth}{0.86\textheight},center}
\begin{tabulary}{\textwidth}{L C C C C C C}
\toprule
\textbf{Model} & \textbf{Difficulty} & \textbf{EX} & \textbf{Safe-EX} & \textbf{Viol. (\%)} & \textbf{OR (\%)} & \textbf{AC-F1} \\
\midrule

\multirow{3}{*}{\makecell[l]{Snowflake- \\ R1-7b}}
  & Simple    & 52.74 & 68.03$\pm$0.87 & 59.95$\pm$1.46 & 0.05$\pm$0.06 & 56.85$\pm$1.56 \\
  & Moderate  & 32.73 & 49.75$\pm$1.73 & 66.77$\pm$1.02 & 0.14$\pm$0.18 & 49.65$\pm$1.30 \\
  & Challenge & 19.05 & 47.01$\pm$6.00 & 81.04$\pm$1.21 & 0.00$\pm$0.00 & 31.65$\pm$1.93 \\
\cmidrule(lr){2-7}

\multirow{3}{*}{\makecell[l]{Llama3- \\ SQLCoder- \\ 8b}}
  & Simple    & 32.13 & 47.61$\pm$2.03 & 61.23$\pm$1.59 & 0.00$\pm$0.00 & 55.79$\pm$1.69 \\
  & Moderate  & 10.38 & 21.39$\pm$3.81 & 67.95$\pm$1.04 & 0.00$\pm$0.00 & 48.54$\pm$1.20 \\
  & Challenge & 6.06 & 21.39$\pm$3.00 & 81.21$\pm$1.05 & 0.00$\pm$0.00 & 31.62$\pm$1.48 \\
\cmidrule(lr){2-7}

\multirow{3}{*}{\makecell[l]{Gemma3-4b}}
  & Simple    & 32.01 & 45.27$\pm$1.73 & 61.86$\pm$1.80 & 0.00$\pm$0.00 & 55.12$\pm$1.87 \\
  & Moderate  & 11.74 & 23.64$\pm$3.55 & 67.27$\pm$0.82 & 0.00$\pm$0.00 & 49.32$\pm$0.93 \\
  & Challenge & 5.63 & 16.92$\pm$2.96 & 81.73$\pm$1.29 & 0.00$\pm$0.00 & 30.87$\pm$1.87 \\
\cmidrule(lr){2-7}

\multirow{3}{*}{\makecell[l]{Gemma3- \\ 27b}}
  & Simple    & 53.20 & 58.64$\pm$1.04 & 30.90$\pm$1.76 & 1.91$\pm$0.45 & 70.06$\pm$2.01 \\
  & Moderate  & 29.80 & 34.89$\pm$4.27 & 36.79$\pm$1.84 & 1.76$\pm$0.26 & 62.29$\pm$1.50 \\
  & Challenge & 15.15 & 32.51$\pm$5.59 & 36.36$\pm$1.32 & 0.52$\pm$0.51 & 50.45$\pm$1.64 \\
\cmidrule(lr){2-7}

\multirow{3}{*}{\makecell[l]{Qwen2.5- \\ Coder-7b- \\ Instruct}}
  & Simple    & 43.66 & 54.40$\pm$1.38 & 60.82$\pm$1.60 & 0.00$\pm$0.00 & 56.24$\pm$1.65 \\
  & Moderate  & 19.64 & 29.49$\pm$3.94 & 67.90$\pm$1.12 & 0.00$\pm$0.00 & 48.52$\pm$1.29 \\
  & Challenge & 10.82 & 24.08$\pm$4.13 & 81.21$\pm$1.52 & 0.00$\pm$0.00 & 31.51$\pm$2.05 \\
\cmidrule(lr){2-7}

\multirow{3}{*}{\makecell[l]{Qwen2.5- \\ 14b- \\ Instruct}}
  & Simple    & 40.40 & 53.14$\pm$1.07 & 50.71$\pm$1.23 & 0.14$\pm$0.09 & 60.47$\pm$1.48 \\
  & Moderate  & 16.48 & 32.10$\pm$3.06 & 59.19$\pm$1.89 & 0.14$\pm$0.11 & 52.28$\pm$1.59 \\
  & Challenge & 8.66 & 26.65$\pm$2.32 & 70.82$\pm$2.50 & 0.09$\pm$0.17 & 33.78$\pm$2.38 \\
\cmidrule(lr){2-7}

\multirow{3}{*}{\makecell[l]{Deepseek- \\ V3.2- \\ Reasoner}}
  & Simple    & 51.22 & 63.86$\pm$1.10 & 7.92$\pm$0.69 & 1.26$\pm$0.27 & 89.66$\pm$0.40 \\
  & Moderate  & 30.70 & 46.23$\pm$4.60 & 8.94$\pm$0.85 & 1.67$\pm$0.46 & 85.95$\pm$1.74 \\
  & Challenge & 14.72 & 36.21$\pm$5.23 & 11.95$\pm$0.97 & 1.21$\pm$0.50 & 73.28$\pm$3.03 \\
\cmidrule(lr){2-7}

\multirow{3}{*}{\makecell[l]{Deepseek- \\ V3.2- \\ Coder}}
  & Simple    & 49.94 & 61.82$\pm$1.19 & 44.17$\pm$1.33 & 0.19$\pm$0.16 & 64.09$\pm$1.61 \\
  & Moderate  & 28.44 & 47.15$\pm$3.68 & 50.48$\pm$2.11 & 0.09$\pm$0.11 & 56.55$\pm$1.76 \\
  & Challenge & 13.85 & 31.26$\pm$6.64 & 70.13$\pm$1.88 & 0.00$\pm$0.00 & 33.69$\pm$2.48 \\
\cmidrule(lr){2-7}

\multirow{3}{*}{\makecell[l]{Claude- \\ Sonnet-4.5}}
  & Simple    & 65.19 & 70.14$\pm$0.43 & 7.45$\pm$0.45 & 1.91$\pm$0.23 & 89.22$\pm$0.58 \\
  & Moderate  & 44.92 & 57.01$\pm$3.60 & 7.99$\pm$1.04 & 1.81$\pm$0.57 & 86.79$\pm$1.40 \\
  & Challenge & 27.71 & 48.72$\pm$7.58 & 8.05$\pm$1.05 & 1.47$\pm$0.70 & 79.16$\pm$3.68 \\
\cmidrule(lr){2-7}

\multirow{3}{*}{\makecell[l]{Gemini- \\ 2.5-Flash}}
  & Simple    & 61.35 & 72.85$\pm$0.75 & 19.00$\pm$1.24 & 0.95$\pm$0.15 & 79.79$\pm$1.10 \\
  & Moderate  & 44.92 & 57.34$\pm$2.29 & 23.21$\pm$2.69 & 0.95$\pm$0.36 & 73.02$\pm$2.62 \\
  & Challenge & 28.57 & 49.39$\pm$5.73 & 25.72$\pm$2.33 & 0.69$\pm$0.44 & 58.27$\pm$3.14 \\
\cmidrule(lr){2-7}

\multirow{3}{*}{\makecell[l]{GPT-4o- \\ mini}}
  & Simple    & 46.45 & 47.23$\pm$1.16 & 25.59$\pm$1.22 & 6.54$\pm$0.76 & 67.50$\pm$0.96 \\
  & Moderate  & 24.60 & 30.34$\pm$2.39 & 33.72$\pm$1.07 & 3.02$\pm$0.31 & 62.09$\pm$1.39 \\
  & Challenge & 12.55 & 23.57$\pm$0.76 & 36.97$\pm$1.70 & 3.38$\pm$0.57 & 43.72$\pm$2.74 \\
\cmidrule(lr){2-7}

\multirow{3}{*}{\makecell[l]{GPT-5- \\ mini}}
  & Simple    & 53.43 & 63.96$\pm$0.91 & 10.99$\pm$1.04 & 0.79$\pm$0.22 & 87.16$\pm$0.87 \\
  & Moderate  & 31.15 & 44.37$\pm$3.20 & 13.68$\pm$1.54 & 0.41$\pm$0.22 & 82.76$\pm$1.72 \\
  & Challenge & 16.02 & 38.60$\pm$7.02 & 17.06$\pm$1.54 & 0.61$\pm$0.35 & 68.37$\pm$3.36 \\
\cmidrule(lr){2-7}

\multirow{3}{*}{\makecell[l]{GPT-5}}
  & Simple    & 55.88 & 68.60$\pm$1.62 & 9.36$\pm$0.90 & 0.86$\pm$0.09 & 88.73$\pm$0.65 \\
  & Moderate  & 32.51 & 44.39$\pm$5.50 & 11.20$\pm$1.20 & 0.41$\pm$0.26 & 85.39$\pm$1.51 \\
  & Challenge & 15.58 & 36.57$\pm$7.66 & 13.33$\pm$1.04 & 0.52$\pm$0.33 & 73.55$\pm$2.62 \\

\bottomrule
\end{tabulary}
\end{adjustbox}
\label{tab:bird-singlecol}
\end{table*}

\begin{table*}[htbp]
\caption{RBAC-Spider difficulty tier performance.}
\label{tab:spider-singlecol}
\centering
\small
\begin{adjustbox}{max totalsize={\textwidth}{0.86\textheight},center}
\begin{tabulary}{\textwidth}{L C C C C C C}
\toprule
\textbf{Model} & \textbf{Difficulty} & \textbf{EX} & \textbf{Safe-EX} & \textbf{V (\%)} & \textbf{OR (\%)} & \textbf{AC-F1} \\
\midrule

\multirow{4}{*}{\makecell[l]{Snowflake-R1-7b}}
  & Easy   & 89.75 & 86.57$\pm$1.66 & 34.60$\pm$1.83 & 0.24$\pm$0.32 & 78.59$\pm$1.54 \\
  & Medium & 83.50 & 77.90$\pm$1.35 & 51.75$\pm$1.73 & 0.00$\pm$0.00 & 64.45$\pm$1.61 \\
  & Hard   & 74.66 & 67.01$\pm$3.25 & 52.53$\pm$1.84 & 0.11$\pm$0.23 & 63.86$\pm$1.62 \\
  & Extra  & 60.40 & 58.71$\pm$4.25 & 57.47$\pm$3.31 & 0.00$\pm$0.00 & 59.41$\pm$3.11 \\
\cmidrule(lr){2-7}

\multirow{4}{*}{\makecell[l]{Llama3-SQLCoder-8b}}
  & Easy   & 75.00 & 76.15$\pm$2.97 & 35.89$\pm$2.35 & 0.00$\pm$0.00 & 78.11$\pm$1.74 \\
  & Medium & 61.66 & 63.77$\pm$1.71 & 53.63$\pm$1.59 & 0.00$\pm$0.00 & 63.34$\pm$1.47 \\
  & Hard   & 46.55 & 41.71$\pm$2.80 & 54.48$\pm$1.52 & 0.00$\pm$0.00 & 62.54$\pm$1.42 \\
  & Extra  & 39.16 & 36.54$\pm$5.34 & 59.16$\pm$2.89 & 0.00$\pm$0.00 & 57.94$\pm$2.88 \\
\cmidrule(lr){2-7}

\multirow{4}{*}{\makecell[l]{Gemma3-4b}}
  & Easy   & 87.50 & 80.94$\pm$1.20 & 35.48$\pm$2.69 & 0.00$\pm$0.00 & 78.34$\pm$1.95 \\
  & Medium & 72.20 & 73.08$\pm$1.78 & 53.72$\pm$1.64 & 0.00$\pm$0.00 & 63.26$\pm$1.53 \\
  & Hard   & 57.47 & 55.59$\pm$2.80 & 53.68$\pm$2.20 & 0.00$\pm$0.00 & 63.28$\pm$2.04 \\
  & Extra  & 43.37 & 39.82$\pm$4.50 & 59.04$\pm$3.55 & 0.00$\pm$0.00 & 58.03$\pm$3.57 \\
\cmidrule(lr){2-7}

\multirow{4}{*}{\makecell[l]{Gemma3-27b}}
  & Easy   & 93.15 & 87.94$\pm$1.41 & 17.42$\pm$1.79 & 3.07$\pm$0.20 & 86.14$\pm$1.60 \\
  & Medium & 85.20 & 81.23$\pm$1.58 & 36.19$\pm$1.88 & 0.13$\pm$0.11 & 72.07$\pm$1.70 \\
  & Hard   & 71.26 & 66.28$\pm$5.37 & 43.45$\pm$1.88 & 0.69$\pm$0.23 & 67.98$\pm$1.63 \\
  & Extra  & 56.02 & 52.51$\pm$4.96 & 43.86$\pm$4.99 & 0.12$\pm$0.24 & 65.61$\pm$4.17 \\
\cmidrule(lr){2-7}

\multirow{4}{*}{\makecell[l]{Qwen2.5-Coder-7b-Instruct}}
  & Easy   & 91.53 & 85.96$\pm$1.04 & 35.49$\pm$3.27 & 0.00$\pm$0.00 & 78.32$\pm$2.37 \\
  & Medium & 80.94 & 75.57$\pm$0.94 & 52.96$\pm$1.45 & 0.00$\pm$0.00 & 63.88$\pm$1.35 \\
  & Hard   & 66.67 & 63.84$\pm$2.55 & 52.99$\pm$1.28 & 0.00$\pm$0.00 & 63.94$\pm$1.19 \\
  & Extra  & 40.96 & 42.24$\pm$4.79 & 58.43$\pm$2.83 & 0.00$\pm$0.00 & 58.67$\pm$2.80 \\
\cmidrule(lr){2-7}

\multirow{4}{*}{\makecell[l]{Qwen2.5-14b-Instruct}}
  & Easy   & 86.29 & 79.62$\pm$2.85 & 21.53$\pm$2.45 & 1.45$\pm$0.66 & 84.96$\pm$1.98 \\
  & Medium & 76.91 & 73.48$\pm$2.14 & 38.83$\pm$1.89 & 0.45$\pm$0.14 & 70.29$\pm$1.62 \\
  & Hard   & 59.77 & 58.15$\pm$1.30 & 42.64$\pm$3.66 & 0.00$\pm$0.00 & 69.12$\pm$2.39 \\
  & Extra  & 39.16 & 41.64$\pm$5.82 & 48.92$\pm$4.11 & 0.00$\pm$0.00 & 62.53$\pm$3.87 \\
\cmidrule(lr){2-7}

\multirow{4}{*}{\makecell[l]{Deepseek-V3.2-Reasoner}}
  & Easy   & 90.57 & 85.62$\pm$0.65 & 1.13$\pm$0.90 & 3.87$\pm$0.70 & 96.29$\pm$1.15 \\
  & Medium & 82.49 & 80.01$\pm$0.82 & 2.11$\pm$0.50 & 1.62$\pm$0.52 & 96.25$\pm$0.56 \\
  & Hard   & 60.27 & 65.26$\pm$2.95 & 1.84$\pm$0.67 & 2.30$\pm$0.73 & 95.83$\pm$0.73 \\
  & Extra  & 64.80 & 51.36$\pm$8.23 & 3.86$\pm$1.64 & 2.17$\pm$1.35 & 93.18$\pm$2.91 \\
\cmidrule(lr){2-7}

\multirow{4}{*}{\makecell[l]{Deepseek-V3.2-Coder}}
  & Easy   & 90.98 & 82.23$\pm$0.84 & 19.92$\pm$1.74 & 1.53$\pm$0.53 & 85.90$\pm$1.75 \\
  & Medium & 82.49 & 77.73$\pm$1.82 & 34.39$\pm$1.25 & 0.36$\pm$0.11 & 73.00$\pm$1.24 \\
  & Hard   & 65.07 & 53.10$\pm$1.72 & 40.35$\pm$2.58 & 0.34$\pm$0.28 & 70.05$\pm$1.99 \\
  & Extra  & 69.20 & 47.25$\pm$5.88 & 44.82$\pm$3.29 & 0.00$\pm$0.00 & 65.21$\pm$3.35 \\
\cmidrule(lr){2-7}

\multirow{4}{*}{\makecell[l]{Claude-Sonnet-4.5}}
  & Easy   & 91.53 & 88.76$\pm$1.05 & 2.18$\pm$0.83 & 3.79$\pm$0.32 & 95.60$\pm$0.85 \\
  & Medium & 89.91 & 81.13$\pm$2.14 & 5.29$\pm$0.23 & 2.60$\pm$0.59 & 92.17$\pm$0.82 \\
  & Hard   & 82.18 & 79.00$\pm$2.13 & 5.40$\pm$1.06 & 3.22$\pm$0.93 & 91.55$\pm$1.36 \\
  & Extra  & 67.47 & 61.19$\pm$5.54 & 7.35$\pm$1.68 & 2.77$\pm$1.35 & 88.95$\pm$2.97 \\
\cmidrule(lr){2-7}

\multirow{4}{*}{\makecell[l]{Gemini-2.5-Flash}}
  & Easy   & 94.26 & 87.71$\pm$1.61 & 3.06$\pm$0.65 & 3.71$\pm$0.54 & 95.03$\pm$0.92 \\
  & Medium & 89.85 & 81.93$\pm$2.29 & 10.49$\pm$0.48 & 1.88$\pm$0.48 & 88.32$\pm$0.83 \\
  & Hard   & 86.99 & 71.59$\pm$3.37 & 14.60$\pm$0.93 & 1.61$\pm$0.43 & 85.34$\pm$1.05 \\
  & Extra  & 64.00 & 68.30$\pm$7.91 & 19.64$\pm$1.24 & 0.72$\pm$0.59 & 80.87$\pm$0.65 \\
\cmidrule(lr){2-7}

\multirow{4}{*}{\makecell[l]{GPT-4o-mini}}
  & Easy   & 90.98 & 66.71$\pm$1.47 & 5.16$\pm$0.86 & 18.31$\pm$2.05 & 81.14$\pm$2.08 \\
  & Medium & 79.70 & 65.42$\pm$1.87 & 16.10$\pm$1.18 & 8.52$\pm$0.79 & 76.24$\pm$1.12 \\
  & Hard   & 61.64 & 49.17$\pm$2.98 & 24.94$\pm$2.35 & 7.13$\pm$0.59 & 71.94$\pm$2.61 \\
  & Extra  & 58.80 & 41.61$\pm$6.26 & 28.79$\pm$3.56 & 6.87$\pm$1.89 & 66.27$\pm$4.79 \\
\cmidrule(lr){2-7}

\multirow{4}{*}{\makecell[l]{GPT-5-mini}}
  & Easy   & 90.16 & 79.30$\pm$0.93 & 1.29$\pm$0.59 & 3.15$\pm$0.97 & 96.71$\pm$1.15 \\
  & Medium & 77.66 & 73.15$\pm$1.27 & 4.35$\pm$0.69 & 1.12$\pm$0.42 & 94.63$\pm$0.61 \\
  & Hard   & 68.49 & 65.66$\pm$4.47 & 2.30$\pm$1.03 & 1.72$\pm$0.36 & 96.02$\pm$1.17 \\
  & Extra  & 60.40 & 53.38$\pm$5.17 & 6.38$\pm$1.05 & 1.20$\pm$0.76 & 91.79$\pm$1.93 \\
\cmidrule(lr){2-7}

\multirow{4}{*}{\makecell[l]{GPT-5}}
  & Easy   & 89.75 & 74.52$\pm$1.47 & 1.37$\pm$0.41 & 3.63$\pm$0.57 & 96.30$\pm$0.63 \\
  & Medium & 77.41 & 68.58$\pm$1.17 & 3.14$\pm$0.59 & 1.26$\pm$0.54 & 95.63$\pm$0.55 \\
  & Hard   & 68.49 & 62.64$\pm$2.97 & 1.49$\pm$0.46 & 1.95$\pm$0.46 & 96.55$\pm$0.65 \\
  & Extra  & 58.00 & 54.44$\pm$7.63 & 4.22$\pm$1.26 & 1.45$\pm$0.98 & 93.74$\pm$2.34 \\

\bottomrule
\end{tabulary}
\end{adjustbox}
\end{table*}


\end{document}